\documentclass[a4paper,fleqn,usenatbib]{mnras}
\usepackage{deluxetable}
\usepackage{rotating}
\usepackage{lscape}
\usepackage{graphicx}
\usepackage{subfigure}
\usepackage{amsmath}
\usepackage{amssymb}
\usepackage{IEEEtrantools}

\usepackage[T1]{fontenc}
\usepackage{ae,aecompl}

\title[Magnetic field in ETGs]{Far-infrared - radio correlation and magnetic field strength in star-forming early-type galaxies}

\author[Omar \& Paswan]{
A. Omar,$^{1}$\thanks{E-mail: aomar@aries.res.in}
A. Paswan$^{1,2}$
\\
$^{1}$Aryabhatta Research Institute of observational sciences, Manora Peak, Nainital, India\\
$^{2}$Pt. Ravishankar Shukla University, Raipur, India}

\date{Accepted XXX. Received YYY; in original form ZZZ}

\pubyear{2017}

\begin{document}
\label{firstpage}
\pagerange{\pageref{firstpage}--\pageref{lastpage}}
\maketitle

\begin{abstract}

A tight far-infrared - radio correlation similar to that in star-forming late-type galaxies is also indicated in star-forming blue early-type galaxies, in which the nuclear optical-line emissions are primarily due to star-forming activities without a significant contribution from active galactic nuclei.  The average value of far-infrared to 1.4 GHz radio flux-ratio commonly represented as the $'q'$ parameter is estimated to be $2.35\pm0.05$ with a scatter of 0.16 dex. The average star formation rate estimated using 1.4 GHz radio continuum is $\sim6$ M$_{\odot}$~yr$^{-1}$ in good agreement with those estimated using far-infrared and H$\alpha$ luminosities. The radio emission is detected mainly from central region which could be associated with the star-forming activities, most likely triggered by recent tidal interactions. The average thermal contribution to total radio flux is estimated to be $\sim12$ per cent. The average value of the magnetic field strengths in the star-forming early-type galaxies is estimated to be 12$^{+11}_{-4}$~$\mu$G. These magnetic fields are very likely generated via fast amplification in small-scale turbulent dynamos acting in the star-bursting regions. 
\end{abstract}




\section{Introduction} 

A tight correlation between radio-continuum (RC) emission and infrared (IR) emission from the nuclei of late-type star-forming (SF) galaxies is known since the early studies of van der Kruit (1971, 1973). Later, a far-infrared - radio (FIR-RC) correlation was confirmed in the disks of different types of galaxies such as optical and infrared bright galaxies (Condon, Anderson \& Helou 1991; Yun, Reddy \& Condon 2001; hereafter Y01), dwarf irregular galaxies (e.g., Roychowdhury \& Chengalur 2012), Seyferts and low-ionization nuclear emission regions - LINERs (e.g., Mori\'c et al. 2010), and Wolf-Rayet galaxies with nascent star-formation (e.g., Jaiswal \& Omar 2016).  A large number of studies on the FIR-RC correlations based on observations and theoretical modeling have been carried out. Almost all studies on SF galaxies indicate a tight FIR-RC correlation. Most of the FIR-RC correlations are constructed using total radio flux at cm-wavelengths, where galaxies usually have significant thermal radio emission (free-free) mixed with the dominating non-thermal (synchrotron) radio emission. Tight FIR-RC correlations have also been seen at meter-wavelengths (Basu et al. 2012) where the thermal radio emission is negligible, and also in faint dwarf irregular galaxies where the thermal radio emission can be almost equal to the non-thermal radio component (Roychowdhury \& Chengalur 2012).

The FIR-RC correlation and its tightness are one of the unsolved mysteries in astronomy. The tight FIR-RC correlation is most often linked to the star-forming activities in galaxies. Harwit \& Pacini (1975) proposed that both RC and FIR emission may be linked to massive stars in SF galaxies.  According to a widely discussed scenario (V\"olk 1989; Xu 1990), ultra-violet photons from the massive ($M\ge$8~M$_\odot$) supernovae progenitor stars are considered as the main source for heating of dust in the inter-stellar medium (ISM), which in turn emits in IR wavebands. The same massive star population, which is also short-lived ($\le$10~Myr), undergo type-II supernova explosions in which the cosmic-ray (CR) particles are accelerated to the relativistic energies. These high-energy CR particles loose energy mainly via the synchrotron process in presence of galactic magnetic fields. This scenario requires a fine balance to be maintained between the complex processes of generating CR particles, escaping of CR electrons, synchrotron cooling time, magnetic field strengths and heating of dust so that a universal FIR-RC correlation can be obtained. Such a requirement appears almost like a 'conspiracy' and explanations for such a balance in galaxies with widely varying luminosities over several orders of magnitude are challenging. Nevertheless, several explanations highlighting the possibilities of inter-connections between star-formation, dust heating, cosmic-rays, and magnetic field to understand this apparent conspiracy have been given (Helou \& Bicay 1993; Lisenfeld, V\"olk \& Xu  1996; Nicklas \& Beck 1997; Thompson et al. 2006; Lacki, Thompson \& Quataert 2010). The FIR-RC correlation is also explained via amplification of random turbulent magnetic fields through supernovae explosions in the ISM (Schleicher \& Beck 2013). Some alternate explanations such as the CR heating of molecular clouds and regulation of FIR and radio emission by hydrostatic pressure have also been discussed (Suchkov, Allen \& Heckman 1993; Murgia et al. 2005). The powerful active galactic nuclei (AGN) also display a FIR-RC correlation, however, the relations in AGNs are different and not very tight (Bally \& Thronson 1989; Condon et al. 1991;  Wrobel \& Heeschen 1991; Colina \& P\'erez-Olea 1995; Miller \& Owen 2001). The tightness of the FIR-RC correlation in the SF galaxies has often been used to distinguish the AGN hosting SF galaxies from the SF-dominant galaxies (e.g., Y01; Omar \& Dwarakanath 2005).  

Studies on the FIR-RC correlation in SF early-type galaxies (ETG; lenticulars and ellipticals) are rare although a presence of ISM and star-formation in a large number of ETGs is known since the detection of dust in ETGs via the far-infrared ($25\mu$m$-100\mu$m) emission from the Infrared Astronomical Satellite ($IRAS$; Neugebauer et al. 1984; Jura 1986). The prime reason for this missing study in ETGs is the lack of strong star formation and a high possibility for the presence of AGN in these galaxies. Therefore, even if low-level SF activities were previously detected in a few ETGs, it was difficult to separate emissions from the ISM with that related to an AGN (Bally \& Thronson 1989; Walsh et al. 1989; Wrobel \& Heeschen 1991). More recently, there are some studies on the FIR-RC correlation in nearby ETGs (Combes et al. 2007; Crocker et al. 2011; Nyland et al. 2017). These studies reported larger scatter in the correlation in ETGs as compared to that in the late-type galaxies. Nyland et al. (2017) reported FIR-excess in many ETGs, selected from a sample of nearby ETGs in ATLAS$^{\rm{3D}}$ project (Cappellari et al. 2011). These ETGs have low star formation rates (SFR) with a median value at nearly $\sim0.15$~M$_{\odot}$~yr$^{-1}$ (Davis et al. 2014). A study of the FIR-RC correlation in five ETGs with high SFR ($1 - 5$~M$_{\odot}$~yr$^{-1}$) by Lucero \& Young (2007) indicated a normal relation similar to that known in the late-type galaxies. Recently, Paswan \& Omar (2016) found three ETGs with radio power $L_{\rm 1.4GHz}>10^{21}$~W~Hz$^{-1}$ having a normal FIR-RC relation. Since a cool 'cirrus' component related to low-mass/evolved stars is known to contribute to the total IR emission in galaxies (Helou 1986), the FIR emission from galaxies with very low SFR is not expected to be tightly linked to the diffuse radio emission, related to massive star formation. The cirrus component can also introduce a non-linearity in the FIR-RC correlation towards low FIR luminosity (Y01). As there is already a hint that the FIR-RC correlation in ETGs with high SFR may be similar to that in the late-type galaxies, such studies need to be expanded to a larger sample of galaxies.

The synchrotron radio detection from the SF galaxies can also be used to infer galactic magnetic fields (Beck \& Krause 2005). All star-forming late-type galaxies without any exception are detected with synchrotron radio emission (Condon 1992). This detection establishes magnetic field in almost all late-type SF galaxies. The strengths of the magnetic fields derived using the synchrotron radio emission range from a few $\mu$G in galaxies with low SFR up to 100~$\mu$G in extreme star-burst galaxies (see Beck et al. 1996; Beck 2016). The typical strengths are around 10~$\mu$G in normal SF spiral galaxies (e.g., Fitt \& Alexander 1993). Due to lack of strong star formation, studies of magnetic fields based on synchrotron radio detection are rare in ETGs. Nyland et al. (2017) made an attempt to constrain magnetic fields in ETGs, however, their sample has low SFR and could also be significantly contaminated by weak AGNs. The best examples close to the early-type morphologies are two early-type disk galaxies NGC 4736 (SAab) (Chy\.zy \& Buta 2008) and NGC 4594 (SAa) (Krause, Wielebinski \& Dumke 2006), in which both regular and random magnetic fields have been detected.

In this paper, we present a study based on the FIR and RC emissions from a sample of ETGs with high SFR. The galaxies are selected from the Sloan Digital Sky Survey (SDSS) based galaxy-zoo project in which a unique population of nearby ($z\sim0.02 - 0.05$) blue star-forming ETGs without AGN was found (Schawinski et al. 2009; hereafter S09). The average SFR is inferred as $\sim5$~M$_{\odot}$~yr$^{-1}$ from an analysis of the H$\alpha$ emission from these ETGs. The nature of the optical emission-lines from the central region was inferred as due to star formation, based on the SDSS optical spectroscopy data. The majority of these galaxies are bulge-dominated ETGs with S\'ersic index in the range 2 to 8 (George 2017). The low redshift of these ETGs made it possible to search for RC emission in NRAO (National Radio Astronomy Observatory) VLA (Very Large Array) Sky Survey (NVSS; Condon et al. 1998) and Faint Images of the Radio Sky at Twenty-centimeters (FIRST; Becker, White \& Helfand 1995), and the FIR emission in the $IRAS$ data-base. The estimates for the RC-based SFR and magnetic field are also presented in this paper. The value of the Hubble constant is taken at 68~km~s$^{-1}$~Mpc$^{-1}$ (Planck Collaboration 2016).

\section{Physical properties of the sample} 

A population of blue ETGs with more than 200 galaxies was detected in the VI data release (Adelman-McCarthy et al. 2008) of the SDSS in the redshift range of $0.02 < z < 0.05$ (S09). The SDSS provides broad-band five-color photometry and optical spectroscopy within a $3''$ region in the visible wavelength range for about one-third of the sky using a 2.5-meter optical telescope (Gunn et al. 2006). S09 presented a comprehensive analysis in terms of color, classifications based on the optical-line emissions, star-formation rates based on the H$\alpha$ emission, masses, and environments of these galaxies. A summary of the properties of these galaxies based on the analysis presented in S09 is given here. The $u - r$ optical colors of the blue ETGs are in the range of $1.3 - 2.5$ in comparison to the redder colors ($2.5 - 3.0$) of quiescent ETGs. The nature of the emission from the central ($\sim3''$) region of these galaxies was classified based on two well-established SF-AGN diagnostics schemes (Baldwin, Phillips \& Terlevich 1981; Kauffmann et al. 2003) using the optical-line ratios of [NII] $\lambda$6583/H$\alpha$ $\lambda$6563 and [OIII] $\lambda$5007/H$\beta$ $\lambda$4861. Using these schemes, 55 blue ETGs were identified as star-forming without a significant contribution from AGN. A larger number of 73 blue ETGs were found to have an AGN or a mix of AGN and SF. The identified AGNs were either Seyferts or Low Ionization Nuclear Emission Regions (LINER). The remaining 76 galaxies were termed as weak emission-line (WEL) galaxies in which it was not possible to discern AGN and SF-related emissions due to poor signal-to-noise ratio (SNR) in the optical spectrum. These 55 blue star-forming galaxies without a significant contribution from AGN constitute our sample for the present study.

S09 also found that the SF ETGs have central stellar velocity dispersions ($\sigma$) between 40 and 150 km~s$^{-1}$ and such blue ETGs are nearly absent at $\sigma$ above 200 km~s$^{-1}$. The blue ETGs in this sample are of intermediate stellar masses (a few times $10^{10}$ M$_{\odot}$) without any massive red-sequence early-type system. These blue ETGs are rare and make up nearly 6 per cent of the low-redshift ETG population. These blue ETGs are residing in low galaxy-density environments. The difference in the local density was also visible as a function of color in the sense that the bluer galaxies are found in lower density environments compared to their red counterparts at the same velocity dispersion (S09). These blue ETGs are not residing in the central regions of cluster of galaxies. These properties make it a unique sample of star-forming ETGs, which was not available before. The SFRs were estimated in the range of $0.5 - 21$ M$_{\odot}$~yr$^{-1}$ from the H$\alpha$ line-flux within the central $3''$ region extrapolated to the optical extents of the galaxies (S09). The bluer early-type galaxies in this sample tend to show higher SFR on average (Paswan \& Omar 2016). They also showed a hint that the SFR is linked to the central stellar velocity dispersion in the sense that the SFR decreases in galaxies with $\sigma$ beyond $\sim100$ km~s$^{-1}$. This trend in SFR with $\sigma$ is consistent with some semi-analytical models in which less-massive galaxies with lower $\sigma$ within 80 - 240 km s$^{-1}$ are predicted to experience star formation (e.g., Schawinski, Khochfar \& Kaviraj 2006). Therefore, it appears that the star-forming ETGs in the S09 sample are having unique environmental and physical properties which are allowing these galaxies to harbor intense star-forming activities.

The morphological designation of these galaxies as presented in S09 was somewhat uncertain as it was based on visual classifications. Most of the galaxies were classified as E/S0. Recently, George (2017) presented structural analysis of the SF galaxies in the S09 sample through S\'ersic profile fitting to the optical surface brightness. This analysis provides a quantitative parameter S\'ersic index, which can be better than qualitative visual classification of galaxy morphology. In our radio-detected sample, thirty-two galaxies showed S\'ersic index in the range $2 - 8$ indicating bulge dominance in these systems, and 6 galaxies have the index $<2$ indicating a possibility for the presence of a disc component. The analysis presented by George (2017) also showed that after subtracting a best-fit S\'ersic profile, the residual images of several galaxies show significant deviations from a smooth light distribution. These deviations are identified as signatures of tidal debris in form of tidal tails, shells and asymmetric excess-light. The residual images also show circum-nuclear ring-like structure. These galaxies also follow the tight Kormendy scaling-relation (see Kormendy 1977) between the effective radius and the mean surface brightness at the effective radius. The Kormendy relation is followed by early-type galaxies. This analysis confirms these galaxies as early-types without an ambiguity.

\begin{table*}
\centering
\caption{Main properties, and radio and far-infrared flux densities of blue star-forming early-type galaxies taken from the S09 sample.}
\begin{tabular}{llcccccc}
\hline
S09 & Galaxy & Redshift & $S^{\rm FIRST}_{1.4 \mathrm{GHz}}$  & $S^{\rm NVSS}_{1.4 \mathrm{GHz}}$ & $S_{60\mu \mathrm{m}}$ & $S_{100\mu \mathrm{m}}$ & Size$^{a}$\\
 id &      & $z$  &   [mJy]      & [mJy]   & [Jy] & [Jy]& [arcsec]\\
\hline
8   &J010358+151450 &   0.04176   &     x              &	 $2.3\pm0.4$  & 	$0.20\pm0.07$   &	$<1.3$        &  32 \\
7   &J014143+134032 &   0.04539   &     x              &	 $3.1\pm0.4$  & 	$0.74\pm0.06$   &	$0.94\pm0.18$ &  25 \\
68  &J030126-000425 &   0.04285   &     $1.02\pm0.12$  &	 $<1.5$       &	$<0.1$          &    $<0.8$        &  28 \\
108 &J074723+222041 &   0.04549   &     $<0.5$         &	 $<1.5$       & 	$<0.2$          &	$<0.9$        & 	22 \\
84  &J075420+255133 &   0.04167   &     $<0.5$         &	 $<1.5$       & 	$0.20\pm0.05$   &	$<1.2$        &  25 \\
139 &J075608+172250 &   0.02899   &     $<0.5$         &	 $<1.5$       & 	$0.17\pm0.04$   &    $<0.4$	      & 	37 \\
137 &J075636+184417 &   0.03988   &     $<0.5$         &	 $<1.5$       & 	$<0.24$         &	$<0.7$        &  33 \\
129 &J075912+533326 &   0.03479   &     $2.10\pm0.17$  &	 $4.8\pm0.5$  &  $0.76\pm0.05$   &	$0.90\pm0.13$ &  30 \\
130 &J081020+561226 &   0.04623   &     $<0.5$         &    $<1.5$       & 	$<0.2$          &	$<0.8$        &  23 \\
172 &J081756+470719 &   0.03901   &     $3.30\pm0.15$  &	 $2.6\pm0.5$  &  $0.27\pm0.07$   &	$1.46\pm0.25$ &  34 \\
207 &J082909+524906 &   0.04842   &     $<0.5$         &	 $<1.5$       & 	$<0.1$          &	$<0.5$        & 	22 \\
213 &J084346+313452 &   0.04756   &     $0.73\pm0.13$  &	 $<1.5$       &	$0.25\pm0.04$   &	$0.49\pm0.17$ & 	16 \\
86  &J085311+370806 &   0.04980   &     $2.40\pm0.13$  &	 $4.2\pm0.4$  & 	$0.28\pm0.04$   &	$1.30\pm0.20$ &  23 \\
175 &J092429+534137 &   0.04590   &     $<0.5$         &	 $<1.5$       & 	$0.10\pm0.02$   &	$0.40\pm0.11$ &  20 \\
41  &J101628+033502 &   0.04848   &     $<0.5$         &	 $<1.5$       & 	$0.13\pm0.04$   &	$<0.4$        & 	33 \\
148 &J102034+291410 &   0.04846   &     $<0.5$         &	 $<1.5$       & 	$0.11\pm0.04$   &	$<0.5$ &  20 \\
160 &J102524+272506 &   0.04973   &     $<0.5$         &	 $<1.5$       & 	$<0.1$          &	$<0.5$        &  26 \\
215 &J102654-003229 &   0.03463   &     $2.80\pm0.14$  &	 $2.7\pm0.4$  & 	$0.65\pm0.07$   &	$2.32\pm0.41$ &  29 \\
66  &J105437+553946 &   0.04787   &     $0.50\pm0.12$      &    $<1.5$       & 	$0.21\pm0.03$   &	$0.43\pm0.12$ &  24 \\
2   &J112327-004248 &   0.04084   &     $0.66\pm0.13$  &	 $<1.5$       & 	$0.23\pm0.07$   &	$<1.1$        &  19 \\
149 &J113122+324222 &   0.03368   &     $4.20\pm0.15$  &	 $4.9\pm0.4$  & 	x               &	x             &  40 \\
192 &J115205+455706 &   0.04316   &     $<0.5$         &	 $<1.5$       & 	$0.13\pm0.04$   &	$<0.5$        &  19 \\
180 &J120617+633819 &   0.03974   &     $5.40\pm0.17$      &    $5.0\pm0.4$  & 	$1.29\pm0.04$   &	$2.13\pm0.17$ &  27 \\
23  &J120647+011709 &   0.04124   &     $<0.5$         &	 $<1.5$       & 	$<0.12$         &	$0.50\pm0.10$ & 	28 \\
5   &J120823+000637 &   0.04081   &     $<0.5$         &	 $<1.5$       & 	$0.13\pm0.04$   &	$0.37\pm0.13$ & 	34 \\
92  &J122023+085137 &   0.04894   &     $<0.5$         &	 $<1.5$       & 	$<0.21$         &	$<0.9$        & 	20 \\
77  &J122037+562846 &   0.04381   &     $<0.5$         &	 $<1.5$       & 	$0.11\pm0.04$   &	$0.77\pm0.10$ &  21 \\
14  &J123502+662233 &   0.04684   &     x              &	 $1.9\pm0.4$  & 	$0.49\pm0.04$   &	$0.76\pm0.12$ &  29 \\
177 &J130141+044049 &   0.03836   &     $<0.5$         &	 $<1.5$       & 	$0.31\pm0.07$   &	$<0.7$        &  26 \\
147 &J132620+314159 &   0.04999   &     $<0.5$         &	 $<1.5$       & 	$0.29\pm0.04$   &	$1.18\pm0.15$ & 	28 \\
121 &J134747+111627 &   0.03942   &     $2.80\pm0.13$  &	 $3.2\pm0.5$  &  $0.33\pm0.05$   &	$0.98\pm0.20$ &  30 \\
50  &J135707+051506 &   0.03967   &     $2.60\pm0.13$  &	 $2.6\pm0.3$  & 	$0.46\pm0.04$   &	$0.84\pm0.12$ &  37 \\
101 &J140248+523000 &   0.04361   &     $<0.5$         &	 $<1.5$       & 	$<0.1$          &	$0.37\pm0.10$ &  24 \\
58  &J140656-013541 &   0.02916   &     $4.40\pm0.15$  &	 $6.2\pm0.5$  & 	$0.82\pm0.06$   &	$1.45\pm0.24$ &	40 \\
102 &J140747+523809 &   0.04381   &     $<0.5$         &	 $<1.5$       & 	$0.18\pm0.03$   &	$0.51\pm0.10$ &  25 \\
190 &J141433+404522 &   0.04185   &     $1.70\pm0.14$	&    $2.5\pm0.4$  & 	$0.37\pm0.04$   &	$0.58\pm0.09$ &  26 \\
202 &J141732+362019 &   0.04712   &     $<0.5$         &	 $<1.5$       & 	$0.20\pm0.04$   &	$<0.3$        & 	21 \\
182 &J143222+565108 &   0.04302   &     $0.65\pm0.12$	&    $3.0\pm0.5$  &	$0.36\pm0.03$   &	$0.80\pm0.14$ &  30 \\
206 &J143733+080443 &   0.04987   &     $<0.5$         &	 $<1.5$       & 	$0.12\pm0.04$   &	$<0.45$       & 	34 \\
52  &J145115+620014 &   0.04306   &     $0.79\pm0.12$  &	 $2.7\pm0.4$  &	$0.36\pm0.02$   &	$1.00\pm0.05$ &  26 \\
195 &J145323+390413 &   0.03153   &     $<0.5$         &	 $<1.5$       & 	$0.14\pm0.03$   &	$0.40\pm0.08$ & 	32 \\
30  &J151719+031918 &   0.03749   &     $<0.5$         &	 $<1.5$       & 	$<0.1$          &	$<0.4$        & 	25 \\
4   &J152347-003823 &   0.03747   &     $0.80\pm0.13$  &	 $<1.5$       & 	$0.18\pm0.06$   &	$<1.2$        & 	25 \\
151 &J154451+175122 &   0.04521   &     $<0.5$         &	 $<1.5$       & 	$0.15\pm0.03$   &	$<0.5$ & 	26 \\
103 &J155000+415811 &   0.03391   &     $1.25\pm0.14$  &	 $2.8\pm0.4$  &	$0.22\pm0.03$   &	$0.65\pm0.09$ &  28 \\
119 &J155335+321820 &   0.04985   &     $<0.5$         &	 $<1.5$       &	$0.13\pm0.04$   &	$<0.2$        & 	23 \\
157 &J160439+164443 &   0.04599   &     $<0.5$         &	 $<1.5$       &  $<0.2$          &	$<0.6$        &  25 \\
146 &J160754+200303 &   0.03165   &     $0.94\pm0.14$  &	 $1.4\pm0.4$  &	$0.31\pm0.04$   &	$0.62\pm0.13$ & 	29 \\ 
209 &J161818+340640 &   0.04733   &     $<0.5$         &	 $<1.5$       &  $<0.1$          &	$0.31\pm0.09$ &  30 \\ 
124 &J164430+195626 &   0.02300   &     $3.90\pm0.14$  &	 $4.80\pm0.4$ & 	$0.69\pm0.05$   &	$0.93\pm0.17$ & 	35 \\
105 &J165116+280652 &   0.04724   &     $1.90\pm0.12$  &	 $2.20\pm0.4$ & 	$0.19\pm0.03$   &	$<2.3$        &  21 \\
56  &J172324+274846 &   0.04845   &     x              &	 $3.50\pm0.5$ & 	$0.25\pm0.05$   &	$0.87\pm0.23$ &  25 \\
61  &J221516-091547 &   0.03843   &     $2.90\pm0.13$  &	 $3.00\pm0.4$ & 	$0.51\pm0.05$   &	$<0.5$        &  41 \\
\hline
\multicolumn{8}{@{}l}{Notes: (1) $^{a}$The sizes are the major axis diameter, as published in the NASA Extra-galactic Database-}\\
\multicolumn{8}{@{}l}{(NED; https://ned.ipac.caltech.edu/) using Two Micron All Sky Survey (2MASS) and SDSS photometry.}\\
\multicolumn{8}{@{}l}{(2) A 'x' value means no data was available for the given location.}\\
\label{table:tab1}
\end{tabular}
\end{table*}

\begin{figure*}
\centering
\includegraphics[width=7.5cm,trim={2cm 3.2cm 3.8cm 3.3cm},clip]{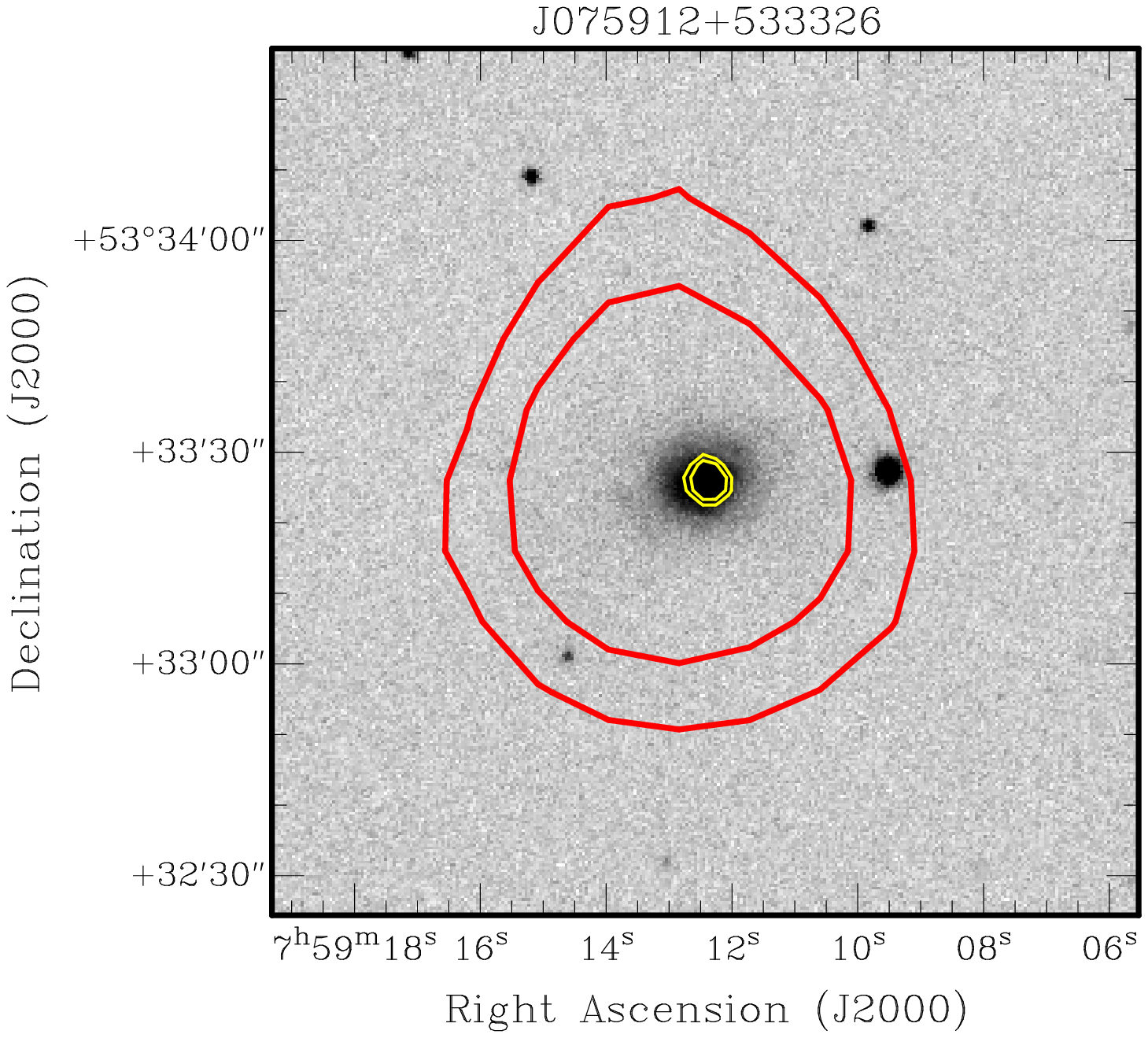}
\includegraphics[width=7.5cm,trim={2cm 3.2cm 3.8cm 3.3cm},clip]{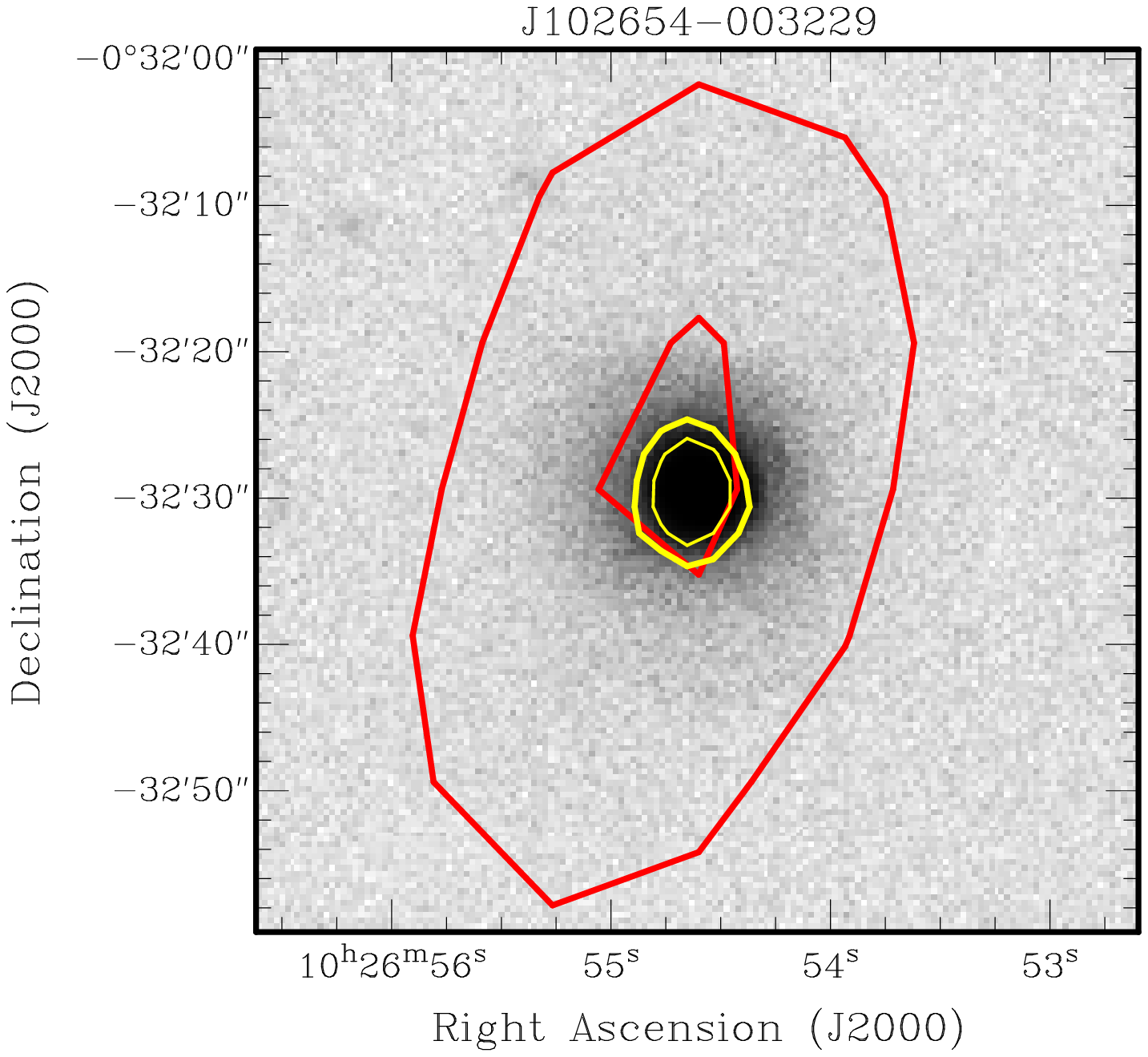}
\caption{The 1.4 GHz radio contours at $3\times$rms and $5\times$rms levels plotted over gray-scale SDSS g-band images of two radio-detected star-forming early-type galaxies. The dark (red) contours are for the NVSS images and the light (yellow) contours are for the FIRST images. Typical values of the rms are $0.15\pm0.02$ mJy and $0.45\pm0.05$ mJy for the FIRST and NVSS contours respectively. The images for the remaining galaxies are provided in the Appendix.}
\label{figure:fig1} 
\end{figure*}

\section{Radio and far-infrared properties of the sample} 

The radio emission at 1.4 GHz was searched in 55 star-forming ETGs using the NVSS and FIRST images. These images were made using the aperture-synthesis interferometric technique (Ryle \& Hewish 1960). The NVSS is a low angular resolution ($45''$ beam) survey with typical rms of $\sim0.45$ mJy~beam$^{-1}$ in the images while the FIRST is a relatively higher angular resolution ($5''$ beam) survey with a typical rms of $\sim0.15$ mJy beam$^{-1}$. The radio images were visually examined for possible detections at the optical locations of the galaxies. Four locations have no observations in the FIRST survey and two source locations appeared confused with a strong background radio source in the vicinity. This reduces our sample to 53 galaxies. The rms (noise) variations among the images or within the images were normally within $\pm0.02$ mJy~beam$^{-1}$ for the FIRST images, and $\pm0.05$ mJy~beam$^{-1}$ for the NVSS images. We detected radio emission in 22 galaxies in the FIRST and 21 galaxies in the NVSS with a total of 17 common detections in FIRST and NVSS. The radio detections are listed in Table~\ref{table:tab1}. The radio images of two galaxies are presented in Fig.~\ref{figure:fig1} and the remaining images are provided in the Appendix. The upper limits to the radio flux density for the undetected galaxies are taken as 0.5 mJy and 1.5 mJy in the FIRST and NVSS images respectively. The SNRs in these detections are above 5, except in two cases where the SNR is nearly 3. Although the radio detections reported here do not have very high SNR, the detections are made in two independent surveys and therefore are highly reliable. 
	
Among the common detections made in the FIRST and NVSS images, it was found that the radio flux densities of the sources estimated from the two surveys differ slightly. In the interferometric aperture synthesis imaging technique, the sensitivity to detect extended radio emission depends on the source size and the density of measurements at short interferometric baselines. Due to low-resolution ($45''$) and high density of measurements taken at short baselines, the NVSS images are not expected to loose any significant flux provided the radio extents of the sources are similar to the optical extents (up to $40''$) of the galaxies in our sample. Becker et al. (1995) showed that the FIRST images can reliably detect sources up to $\sim30''$ and $\sim9''$ extent for total flux greater than nearly 30 mJy and 3 mJy respectively. The radio flux densities of the sources detected here are less than 6 mJy. Therefore, it is possible that some sources (S09 ID: 129, 86, 58, 182, 52, 103), for which the flux densities estimated from the FIRST images are substantially lower compared to that estimated using the NVSS images, are extended beyond $10''$ extent. The maximum difference is $\sim2.7$ mJy among these sources. We also noticed that some sources (S09 ID: 172, 215, 180) have flux density estimated from the FIRST image slightly higher than that estimated using the NVSS image. The differences in all the cases are less than $\sim0.7$ mJy. When comparing the point-like radio sources in the FIRST and the NVSS images, White et al. (1997) found that the difference in flux density estimated from the two surveys is a function of flux density. According to their analysis, the sources with flux densities above $\sim10$ mJy show little scatter, however, weaker sources ($<5$ mJy) can show difference up to $\pm1$ mJy. This difference arises mainly due to noise, flux estimation errors and deconvolution errors. We do not find any source for which the flux density estimated from the FIRST image was substantially ($>1$ mJy) higher compared to that estimated using the NVSS image. Therefore, the minor differences ($<0.7$ mJy) in the flux densities estimated from the two surveys can be due to estimation errors and not due to any intrinsic variation in the source flux.

The far-infrared emissions at 60 $\mu$m and 100 $\mu$m were searched using the online $IRAS$ Scan Processing and Integration (\textsc{scanpi}) tool, which performs 1-dimensional averaging of all the $IRAS$ scans taken for a location in the sky (Helou et al. 1988). The $IRAS$ revised bright galaxy sample is constructed using an earlier version of this tool (Sanders et al. 2003). The \textsc{scanpi} processing tool is known to provide a factor of 2 to 5 gain in the sensitivity over that in the $IRAS$ point source catalogue. The {\textsc{scanpi}} has been reliably used in some studies to get FIR fluxes for weak extended sources (e.g., Lisenfeld et al. 2007). The effective resolution of the $IRAS$ data products is $\sim2'$. We extracted the peak flux densities from the noise-weighted mean of all the $IRAS$ scans using the \textsc{scanpi} tool. Typical rms were nearly 0.04 Jy and 0.12 Jy at 60 $\mu$m and 100 $\mu$m respectively. The detections using the \textsc{scanpi} are made here above 0.1 Jy and 0.3 Jy at 60$\mu$m and 100$\mu$m respectively. In comparison, the $IRAS$ faint source catalogue has sources above 0.2 Jy and 1.0 Jy at 60$\mu$m and 100$\mu$m respectively. The detections with 60 $\mu$m and 100 $\mu$m flux densities are listed in Table~\ref{table:tab1}. The upper limits to the flux for the undetected galaxies are taken as 2.5 times the local rms in the \textsc{scanpi} co-added scan.

\section{Analysis and Results}

\subsection{Far-infrared radio correlation}

The flux densities at the FIR and radio bands were converted to luminosities using the standard relations A1 and A2 given in the Appendix~A. Out of 53 galaxies listed in Table~\ref{table:tab1}, a total of 32 galaxies with S\'ersic index $\ge2$ and 6 galaxies with S\'ersic index $<2$ have radio detections, for which the RC and FIR luminosities are listed in Table~\ref{table:tab2}. Since the galaxies with S\'ersic index $<2$ are likely to have a disk, we do not consider such galaxies as early-type. The SF ETGs with S\'ersic index $\ge2$ have $\langle \log$ ($L_{\rm 1.4 GHz}$/W~Hz$^{-1}\rm )\rangle$  $\sim22.02\pm0.08$. The FIR luminosity $\langle \log$ ($L_{\rm FIR}$/$L_{\odot}\rm ) \rangle$ for the SF ETGs with S\'ersic index $\ge2$ is $\sim10.33\pm0.07$ and $\sim10.22\pm0.07$ for the radio-detected and the FIR detected sample respectively. The upper end ($L_{\rm 1.4 GHz}\sim10^{23}$ W Hz$^{-1}$) of the radio luminosity range of the galaxies in this sample is consistent with that seen in the star-forming late-type galaxies in groups of galaxies (e.g. Omar \& Dwarakanath 2005) and also in nearby clusters of galaxies (e.g., Reddy \& Yun 2004). The ratio of total FIR to radio flux density (or luminosity), commonly represented by a logarithmic $'q'$ parameter (see equations A3 and A4 in the Appendix~A) is plotted in Fig.~\ref{figure:fig3} against the FIR luminosity and the radio luminosity. The reference average value for the $'q'$ parameter is taken as 2.34 from Y01, along with two values at 1.86 and 2.82, indicating a deviation (excess or deficiency) of about three times in radio or FIR luminosity. Our choice of three-times deviation is somewhat stricter than the five-times deviation criterion used in Y01 to identify normal SF late-type galaxies without an AGN component. We used only the NVSS flux density for estimating the $'q'$ parameter as both $IRAS$ and NVSS surveys are sensitive to detect emissions to the full optical extent of the galaxies studied here. Fig. 2 indicates that all the SF ETGs are within the limits of three-times deviation in FIR or RC luminosity. The AGNs are normally identified as radio-excess galaxies with $'q'$ values significantly smaller than the average (Y01, Omar \& Dwarakanath 2005). We do not find any radio-excess galaxies in our sample of SF ETGs. 

The FIR-RC correlations are known to be slightly non-linear (e.g., Nikalas \& Beck 1997; Y01, Mori\'c et al. 2010; Schleicher \& Beck 2013) at IR luminosities below $\sim10^{9}$ L$_{\odot}$ and radio luminosities above $\sim10^{23}$ W Hz$^{-1}$. This non-linearity can be checked from Fig. 2. For a non-linear relation, $'q'$ will not be a constant and depend upon radio or FIR luminosity. For our sample with limited luminosity range, $'q'$ is consistent with being constant. These plots also indicate that the galaxies studied here are not falling within that range of luminosities, where non-linearities in the FIR-RC correlations are prominent and were reported previously. The scatter (rms of residuals) in the $'q'$ parameter for the SF ETGs is estimated to be 0.16 dex, which is similar to that known for the late-type galaxies. The average value of the $'q'$ parameter for the SF ETGs is estimated here at 2.35 $\pm$ 0.05, which is very close to the known average value of 2.34 $\pm$ 0.01 for a large sample of SF late-type galaxies (Y01). This is a strong indication that a tight FIR-RC correlation also exists in the SF early-type galaxies.

\begin{figure*}
\includegraphics[width=6cm,angle=-90]{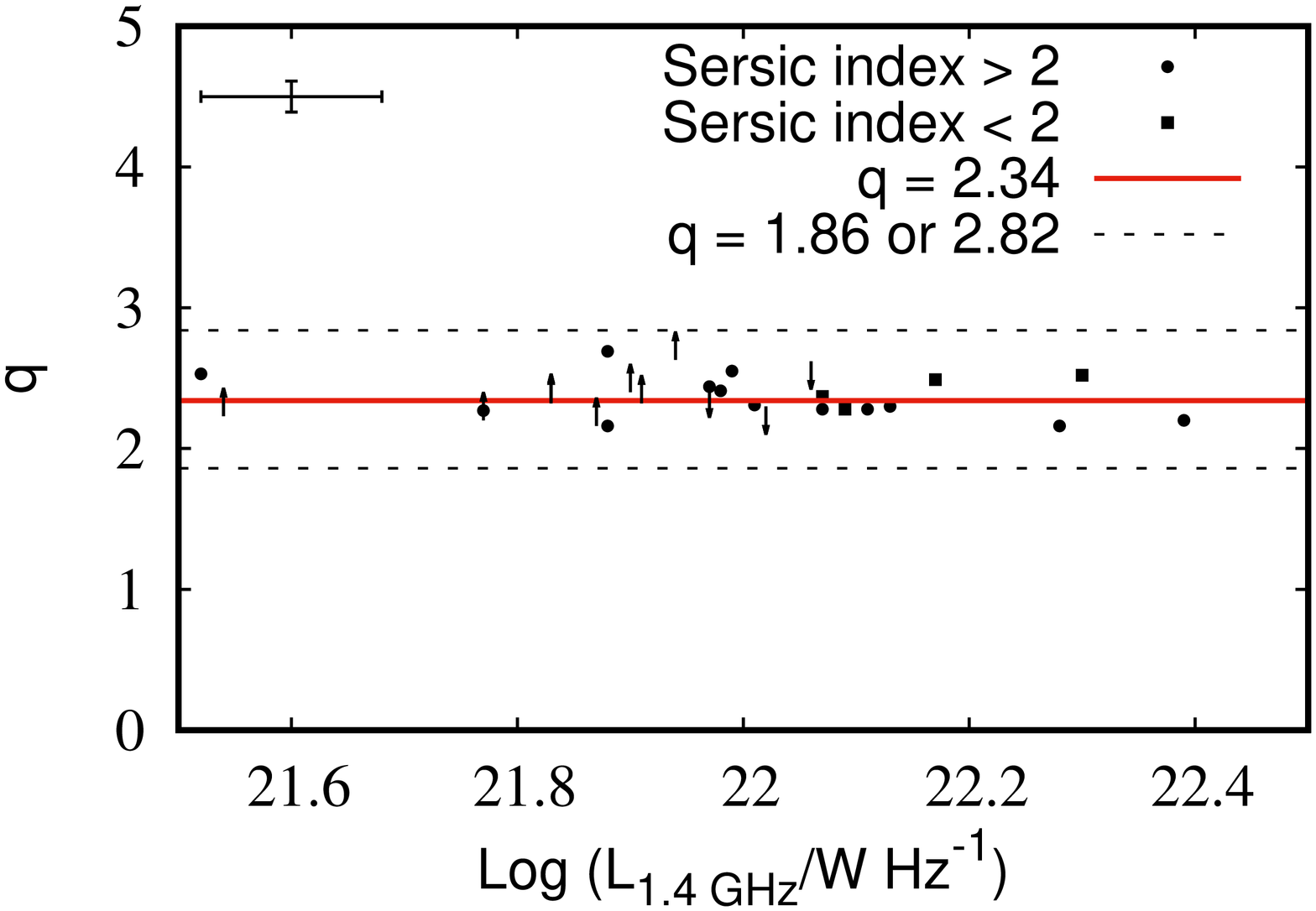}
\includegraphics[width=6cm,angle=-90]{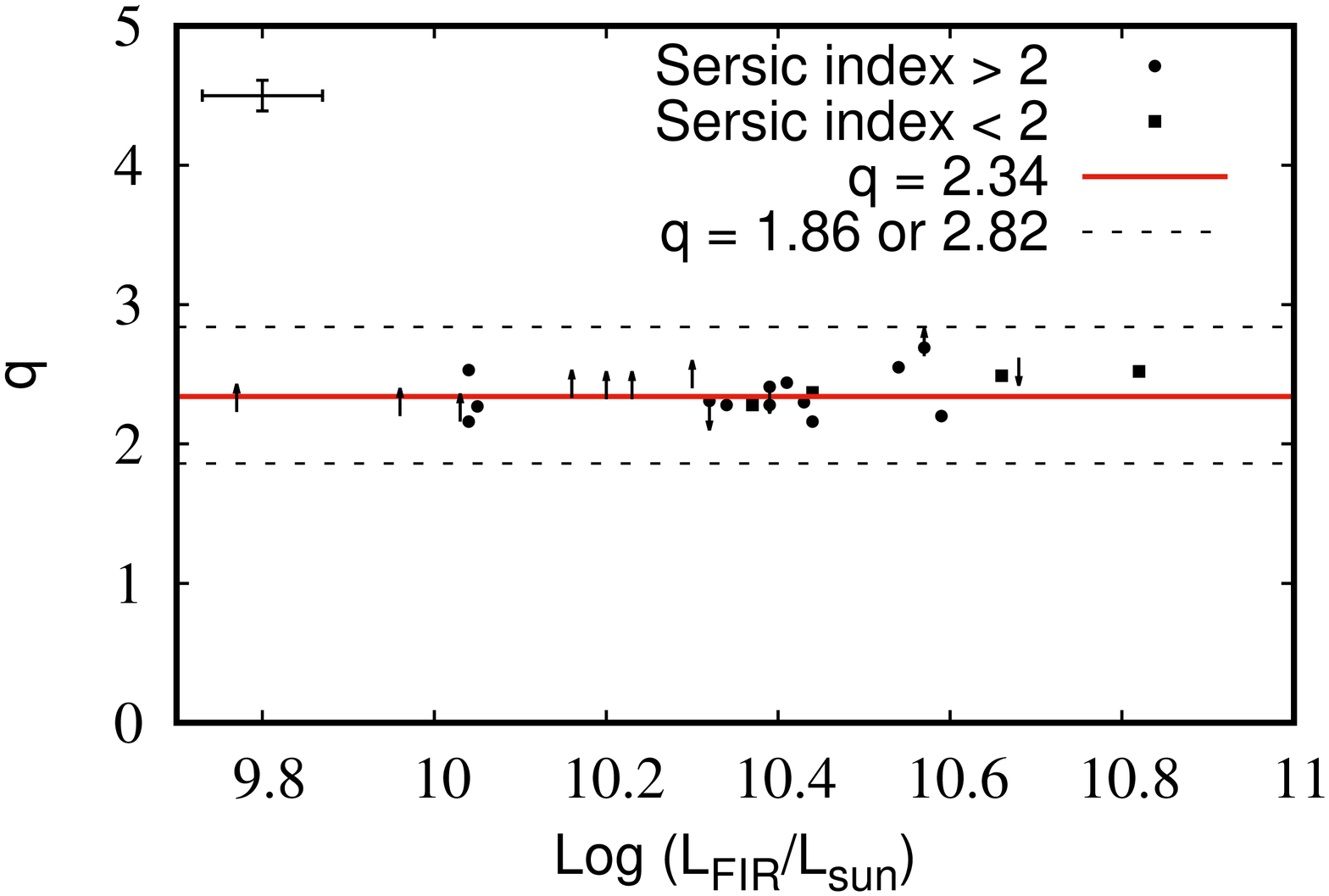}
\caption{The $'q'$ parameter (see equations A3 and A4) for the star-forming ETGs is plotted as a function of radio luminosity (left) and FIR luminosity (right). The error-bars represent typical measurement errors of 0.11, 0.08, 0.07 dex for $'q'$, $L_{\rm 1.4 GHz}$ and $L_{\rm FIR}$ respectively. The solid line represents the average value of the $'q'$ parameter obtained in Y01 for the late-type galaxies. Two dashed lines indicate $'q'$ parameters corresponding to three times deviation (excess or deficiency) in FIR or RC luminosity.}
\label{figure:fig3}
\end{figure*}

\subsection{Nature of radio emission}

The FIRST images show centrally concentrated radio emission in all the galaxies in our sample. The angular resolution of $5''$ in the FIRST images corresponds to nearly 2~kpc and 5~kpc of physical scales at the redshift of 0.02 and 0.05 respectively. Therefore, the FIRST images are sampling a large circum-nuclear region in these galaxies. It is also interesting to note that the residual optical images after a S\'ersic profile fit to the light distribution in these galaxies show a circum-nuclear ring-like morphology of typical extent less than $10''$ (George 2017). This indicates that the star formation in these galaxies is most likely taking place in the circum-nuclear rings. Such circum-nuclear SF rings are common in early-type galaxies (Pogge \& Eskridge 1993; Comer\'on et al. 2014; Kostiuk \& Silchenko 2015). Intense far-ultraviolet emission indicating ongoing star-formation has been detected from the rings seen in the Hubble space telescope images of some S0 galaxies (Salim et al. 2012). These SF rings around galaxies are believed to be formed by resonances (Schommer \& Sullivan 1976) due to non-axisymmetric gravitational potential and accretion of gas (Buta \& Combes 1996). Therefore, we believe that the radio emission detected in the FIRST images is most likely associated with the star-forming circum-nuclear regions in these galaxies. Since the resolution ($45''$) in the NVSS images is much lower compared to that in the FIRST images ($5''$) and also compared to the optical sizes ($<41''$) of the galaxies, it is difficult to compare morphology of the NVSS radio emission with that of the FIRST radio emission.

The powerful AGNs usually have their 1.4 GHz radio luminosities above $\sim10^{23}$ W Hz$^{-1}$. Since the galaxies studied here have radio luminosities in the range $10^{21.5} - 10^{22.4}$ W Hz$^{-1}$, large contaminations due to AGN are not expected. The optical line analyses as presented in S09 carefully isolated AGN hosting SF galaxies from the SF-dominant galaxies. The analyses of population fraction and star-formation history of various types (SF, AGN+SF, WEL) of ETGs indicate that the peak of AGN activity starts roughly 0.5 Gyr after the star-burst phase (Schawinski et al. 2007). Paswan \& Omar (2016) analysed colors of the galaxies in the S09 sample studied here, and predicted similar time-scales for AGN trigger, which within $<1$ Gyr was also predicted to suppress star-formation. The star-bursts in interacting disk galaxies do not last more than nearly 50 Myr (Bernl\"ohr 1993; Mihos \&  Hernquist 1994). Therefore, intense SF and peak AGN activities in ETGs appear well separated in time. Despite these isolations, a contamination from a weak AGN in SF galaxies can not be ruled out in absence of higher resolution radio images. Nevertheless, some more arguments in favor of the radio emission being mainly related to the star formation in these galaxies can be given. A radio image stacking analysis of the WEL sample of ETGs drawn from the same S09 sample studied here, showed a residual average 1.4 GHz radio luminosity as $10^{20.3}$ W Hz$^{-1}$ in absence of intense star-formation (Paswan \& Omar 2006). Therefore, AGNs are weak in these galaxies in general.

The similarity in the average value and the scatter of the $'q'$ parameter in these ETGs with those for the late-type galaxies is also a strong indicator that the majority of radio and FIR emission is likely of a non-AGN origin. Mori\'c et al. (2010) found in a sample of SF and AGN galaxies, that the $'q'$ value decreases with increasing radio luminosity. Their analysis showed a weak negative trend between the $'q'$ parameter and the 1.4 GHz radio luminosity in the range $10^{22}$ and $10^{25}$ W Hz$^{-1}$. The lower $'q'$ values at higher radio luminosities are possibly due to increased contribution of AGNs to the radio luminosity. The trend becomes prominent at radio luminosities above $10^{23}$ W Hz$^{-1}$. Ivison et al. (2010) also reported that the $'q'$ parameter starts showing significantly increased contribution in the radio continuum at $L_{\rm 1.4 GHz} \ge 10^{22.7}$ W Hz$^{-1}$. Moreover, Colina et al. (1995) showed that in radio-quite AGNs, pure starburst model can successfully reproduce the observed FIR-RC correlation in majority of galaxies. Since the radio luminosities of the galaxies in our sample are below $10^{22.4}$ W Hz$^{-1}$, we expect that our sample is not contaminated significantly by an AGN-related radio emission. The nearby SF galaxies, in which high-resolution radio images are available, typically $<10$ per cent of the total radio emission comes from a compact un-resolved core indicative of an AGN (Hill et al. 2001). Therefore, we assume a minor contribution at $10$ per cent level to the total radio emission from the AGN-related emission. We have subtracted this component while estimating SFR and magnetic field.

\subsection{Star formation rates}

The star formation rates can be estimated from the radio and FIR luminosities using the standard conversion relations (see equations A5, A6 and A9 in the Appendix). For several cases, a radio detection corresponding to the FIR or the H$\alpha$ detection is missing. We find that the radio non-detections are consistent with the sensitivity limits of the two radio surveys, given a low level of star forming activities in these galaxies indicated by a weaker FIR or H$\alpha$  emission. The average SFR for the SF ETGs with S\'ersic index $\ge2$ estimated from radio, FIR and H$\alpha$ are $\sim5.9$, $\sim6.5$, and $\sim5.3$ M$_{\odot}$~yr$^{-1}$ respectively. The average SFRs for the common sources detected in both radio and H$\alpha$ are nearly identical at the values of 5.9 and 6.1 M$_{\odot}$~yr$^{-1}$ respectively. Therefore, the average SFRs estimated using three indicators (radio, FIR, and H$\alpha$) are in general consistent with each other. It is to be noted that the radio-SFRs are estimated after subtracting 10 per cent flux (i.e., assumed AGN component) from the total radio flux. In majority of cases, differences in the SFRs estimated using luminosities in different wave-bands are well within a factor of two. This difference can arise due to several uncertainties inherent in the calibration factor to convert luminosity into SFR. 

The SFR estimates are sensitive to variation in the initial mass function (IMF) and the star formation history (Calzetti 2012). The FIR emission usually traces star formation up to $\sim100$ Myr, while the H$\alpha$ emission traces more recent ($\sim10$ Myr) star formation. In case of constant star formation over the past 100 Myr period, different SFR tracers are likely to provide similar estimates for the SFR in a galaxy. If IMFs are bottom-heavy (La Barbera, Ferreras \& Vazdekis 2015), less number of the massive stars responsible for the synchrotron radio emission will be formed. Such IMFs can increase FIR to radio luminosity ratio. The FIR emission is also sensitive to additional parameters such as fraction of the ionizing photons absorbed by hydrogen and dust absorption efficiency of the non-ionizing photons from the young stars (e.g., Inoue, Hirashita \& Kamaya 2000). Moreover, total FIR flux from galaxies without an AGN originates in two components, viz., cold cirrus heated by low-mass/evolved stars and warm dust heated by the young massive stars (Helou 1986). The cirrus component is not linked to recent massive star formation whereas the dominating synchrotron radio emission traces massive star formation activities over a few 100 Myr. A presence of weak AGN can also modify luminosities to some extent in different wave-bands. Overall, comparisons between SFRs estimated from different tracers are complicated due to the reasons discussed above and it is often worth to estimate SFRs in a galaxy from all the wavebands. We noticed that SFRs estimates for only three galaxies (S09 IDs - 61,77,175)  deviate significantly (more than a factor of 3) in different wavebands. These deviations may be due to combinations of several possibilities discussed above.

\subsection{Thermal-radio fraction}

The radio emissions from the SF galaxies contain a small thermal component due to the free-free emission from the non-relativistic electrons in the star-forming regions. The thermal radio emission can be estimated from the H$\alpha$ flux density (or luminosity) as both the emissions are linked to the ionized region powered by the massive stars. The H$\alpha$ luminosities were estimated using the H$\alpha$ based SFR provided in S09. Since the H$\alpha$ measurements as provided in S09 were extrapolated values for the full extents of galaxies from the measurements made via $3''$ SDSS fibre values, such values may have large uncertainty. Therefore, we also estimated thermal-radio fraction using FIR emission as the FIR emission also traces star formation and the FIR measurements from the $IRAS$ are un-resolved for the sizes of the galaxies in this sample. The average of the H$\alpha$-based and the FIR-based estimates for the thermal-radio fraction is provided in Table~2. This analysis is reproduced in the Appendix (see equations A5, A6, A7, A8). The average value of the thermal-radio fraction at 1.4 GHz for the SF ETGs with S\'ersic index $\ge2$ is estimated to be $\sim12$ per cent. It may be noted that both FIR and H$\alpha$ emission provided almost same mean value for the thermal-radio fraction in the range of $3 - 33$ \% for these galaxies. The average value of the thermal-radio fraction in SF ETGs in our sample is similar to that obtained for the late-type star-forming galaxies (e.g., Condon 1992; Condon, Cotton \& Broderick 2002). 

\subsection{Magnetic field in star-forming circum-nuclear region}

The magnetic fields in late-type galaxies usually have a regular field component and a turbulent field component (see e.g., Beck 2016). The regular and turbulent field components can be separated with the help of sensitive radio polarization measurements which are not available for our sample. Here, we estimate total magnetic field in the star-forming ETGs based on the derived synchrotron radio flux from the ISM. The synchrotron flux is taken as 0.78 times the observed radio flux in the FIRST images to remove the assumed 10 per cent contribution due to a possibility of weak AGN as discussed in Sect.~4.2 and 12 per cent average contribution from the thermal-radio component as estimated in Sect.~4.4. The magnetic fields can be estimated via an assumption made from one of the three plausible scenarios - (i) An equipartition between the total energy densities of CR particles and that of the magnetic field, (ii) A pressure equilibrium between the energy densities of CR particles and that of the magnetic field, (iii) Minimizing total energy density. Beck \& Krause (2005) discussed these scenarios in detail and provide formulae to estimate magnetic field. 

The methods to estimate magnetic field in the ISM are described in the Appendix (see equations A10, A11). The radio synchrotron surface brightness (mJy arcsec$^{-2}$) and total path length in the emitting medium are needed to estimate magnetic field using equation A11. These values are not available directly in absence of a detailed high-resolution map of the radio emission. As star-formations in the ETGs are most likely taking place in a ring-like structure, we assume a flattened geometry for estimating field strengths. The source size fitting through deconvolution is not expected to be fruitful due to poor SNR in the radio images. The FIRST images are known to recover radio flux up to $\sim9''$ extent for a flux density $\sim3$ mJy (Becker et al. 1995). We used a radio source size of nearly $10''\times10''$ for all the galaxies in our sample. This size is also consistent with the typical extent of the circum-nuclear rings detected in the optical surface brightness analysis presented by George (2017). The thickness of the synchrotron emitting medium at 1.4 GHz was taken as 1 to 2 kpc in the estimates of field strengths made for the late-type galaxies (Fitt \& Alexander 1993). On the other hand, the turbulent field scale-heights and the CR diffusion scale-lengths in the diffuse ISM of galaxies are estimated up to several kpc from the disk (see Beck et al. 1996). We used here the path length as 4 kpc to estimate total magnetic field in absence of high (sub-kpc) angular resolution radio images. Our choice for the path length is somewhere between that used by Fitt \& Alexander (1993) and estimates for the diffuse ISM given in Beck et al. (1996). The non-thermal radio spectral indices for the ETGs studied here are not known. We assumed spectral index as $\sim0.75$ following the observed values of spectral index in the late-type star-forming galaxies (Condon 1992).

The fields were estimated using the formulation provided in Eq. A11. The estimates of the magnetic fields can be highly uncertain due to poor constraints on the geometry of the synchrotron emitting region. Therefore, we also estimated lower and upper bounds to the estimated field strengths assuming a range of parameters. We took the bounds for the source size as $7''$ and $13''$, and those for the path-length as 2 kpc and 6 kpc.  We obtain an error due to these uncertainties (source size, path-length and spectral index) as $+11/-4~\mu$G for the average estimated field at $\sim12~\mu$G for the SF ETGs.  It is to be noted here that some other fixed parameters (e.g., electron to proton ratio at 100, and spectral index as $\sim0.75$) used in the calculation are also uncertain, and their effects are not included here.

\section{Discussions}

The S09 sample of ETGs with high SFR provided an opportunity to study FIR and RC relation in ETGs. We found that the average and scatter of the $'q'$ parameter of the SF ETGs are similar to the SF late type galaxies. This is a strong indication that a linear relation between radio continuum and far-infrared luminosity is obeyed by star-forming ETGs also. Previously, Nyland et al. (2017) reported high FIR to radio ratio (high $'q'$ value) in about 20 per cent galaxies in the ATLAS$^{\mathrm{3D}}$ sample of ETGs, and discussed several radio-deficiency scenarios in ETGs including a possibility of lower magnetic fields. We do not find any large deviation in the $'q'$ parameter for the SF ETGs in the S09 sample. A close inspection of the properties of the galaxies in the S09 and ATLAS$^{\mathrm{3D}}$ samples reveals two major differences. The ATLAS$^{\mathrm{3D}}$ galaxies have low median SFR ($\sim0.15$ M$_{\odot}$ yr$^{-1}$; Davis et al. 2014) and high median stellar velocity dispersion (130 km s$^{-1}$; Cappellari et al. 2013) compared to the S09 sample ($<$SFR$> = 5 - 6$ M$_{\odot}$ yr$^{-1}$ and median stellar velocity dispersion $80$ km s$^{-1}$). The radio luminosities of the ATLAS$^{\mathrm{3D}}$ galaxies are in fact closer to those estimated for the WEL galaxies in the S09 sample (Paswan \& Omar 2016). Our estimates for the magnetic field do not indicate a possibility of weak field in the star-forming ETGs at high SFRs. 

It is also worth to make a comparison of our study with the studies on the FIR-RC correlation on 'blue-cloud' galaxies at intermediate and high redshifts (Sargent et al. 2010; Jarvis et al. 2010; Ivison et al. 2010; Bourne et al. 2011; Basu et al. 2015). The blue-cloud galaxies also show tight FIR-RC correlation, constructed using stacking of radio and IR images. The blue-cloud galaxies are also star-forming galaxies, selected based on their blue colors and contain similar galaxies in terms of stellar mass and SFR, as studied in this paper. However, the optical morphologies of the blue-cloud galaxies are presently unknown. Therefore, the sample of the blue ETGs studied here is unique in the sense that it is morphologically confirmed sample of blue star-forming early-type galaxies selected based on colors and optical spectroscopy.

It is worth to point out some possibilities which may affect a FIR-RC relation in ETGs. The IMF variations in ETGs with high stellar velocity dispersion may affect the FIR-RC correlation as the IMFs are known to be strongly linked to stellar velocity dispersion in galaxies (La Barbera et al. 2015). The IMFs may become bottom-heavy at higher stellar velocity dispersions. A bottom-heavy IMF produces relatively more number of low-mass stars as compared to those expected in the Kroupa IMF (Kroupa 2002) for a fixed total mass of all the stars. A bottom-heavy IMF can therefore substantially alter supernovae rates and hence the radio synchrotron emission. The FIR-RC correlation in massive ETGs may be affected by a bottom-heavy IMF.  A significant bottom-heavy IMF is not expected in the S09 sample of galaxies due to their low stellar velocity dispersion. Therefore, the S09 sample of ETGs has some unique properties, which make these galaxies suitable for having a normal and tight FIR-RC correlation similar to the late-type galaxies. It appears that the radio luminosities ($10^{21.5} - 10^{22.4}$ W Hz$^{-1}$) of the ETGs in the S09 sample are in a range, above which the the AGN contribution to the radio flux becomes more significant and below which the cirrus component contribution to the IR flux becomes more significant.

The FIR-RC relation is also linked to field amplification in turbulence driven by supernova explosions of the massive stars (Tabatabaei et al. 2013; Schleicher \& Beck 2013; Schober, Schleicher \& Klessen 2016). Although the estimates of magnetic field in galaxies can be uncertain due to various unknown parameters and assumptions made, the field strengths estimated in the SF ETGs are not too different from those measured in SF late-type galaxies using nearly identical assumptions and choice of parameters. The current theories explain magnetic fields in spiral galaxies via a 3-stage process of seeding, amplifying and ordering. The seed fields can be generated in the young Universe by various possibilities such as purely primordial (Durrer \& Neronov 2013), cosmological structure formation (Lazar et al. 2009), plasma fluctuations in protogalaxies (Schlickeiser \& Felten 2013), Biermann mechanism in primordial supernovae remnants (Hanayama et al. 2005), or jets from the first black holes (Rees 2005). The cosmological seed fields can be amplified in galaxies to the equipartition field strengths at nearly $10~\mu$G level in $10 - 100$ Myr of time, in highly efficient small-scale dynamos driven by the turbulence generated in the supernova explosions (Balsara et al. 2004; Arshakian et al. 2009; Schober et al. 2012;  Beck et al. 2012; Schleicher \& Beck 2013). The turbulent kinetic energy is converted into the magnetic energy in small-scale dynamos. This amplified field is random. The field can become regular and ordered by the large-scale dynamos acting in the differential rotation of the ISM in spiral galaxies over a few giga-years of time (Beck et al. 2016). Regular fields are not expected here due to lack of ordered motions in elliptical galaxies. However, it may be noted that a possibility for field regularization over a few giga-year of time-scales in some dynamical components such as the rings in ETGs has been discussed (Moss et al. 2016). Large-scale magnetic fields can also be generated via cosmic-ray driven dynamos as discussed in Hanasz et al. (2009). All these scenarios will be worth to examine with new observations. Our estimates for the magnetic fields ($\sim12~\mu$G) in the circum-nuclear SF regions are therefore in good agreement with those predicted from small-scale dynamos acting in turbulence, driven by the type-II supernovae in the star-bursting regions.

The star formation in the blue ETGs is very likely triggered by recent tidal interactions, which could be inferred from the optical images (George 2017). Since the random magnetic fields decay as a power-law in absence of continuous injection of turbulent energy (e.g., Beck et al. 2012), we do not expect quiescent low-mass ETGs to have strong magnetic fields after a few hundred Myr of cessation of a previous episode of star-burst. Strong central starbursts in interacting disk galaxies are predicted to last for durations shorter than $50$ Myr (Bernl\"ohr 1993; Mihos \&  Hernquist 1994). This in turn implies that the observed magnetic fields in the star-forming ETGs are very likely amplified during the present starburst. If turbulent field amplifications were taking significantly longer as compared to the star-formation period, this would have possibly resulted into a large scatter or deviation in FIR-RC relations (e.g., the $'q'$ parameter) in ETGs which was not seen in the present study. The fast field-amplification in turbulent dynamos is also consistent with the observed correlation between the turbulent field strength and the SFR on sub-galactic scales in some other galaxies (e.g., Tabatabaei et al. 2013; Chy\.zy 2008). Our inference is in good agreement with the theories of random field amplification in turbulent dynamos.

\section{Summary and concluding remarks}

We presented here the FIR-RC correlation in terms of the ratio of FIR to RC luminosity ($'q'$ parameter) in a sample of star-forming ETGs in the redshift range of $0.02 - 0.05$, selected from the SDSS. The early-type morphology was identified with the S\'ersic index values equal to or above 2. The star-forming nature was identified using the optical-line diagnostics schemes. These galaxies are characterized by blue colors ($u-r < 2.5$), intermediate stellar masses (a few times $10^{10}$~M$_{\odot}$) and central stellar velocity dispersion between 40 and 150 km~s$^{-1}$. This sample is a unique morphologically confirmed sample of star-forming early-type galaxies, which is not contaminated from powerful central AGN sources. The main conclusions from this study are as follows: \\

\noindent (1) The average value of the $'q'$ parameter for the star-forming ETGs is estimated to be $2.35\pm0.05$ with a scatter of 0.16 dex. These values are similar to those seen in the late-type star-forming galaxies.\\

\noindent (2) The 1.4 GHz radio emission in the VLA FIRST images is detected mainly from the circum-nuclear star-bursting region.  \\

\noindent (3) The thermal radio continuum fraction in the star-forming ETGs is estimated at around 12 per cent, similar to that in the late-type galaxies.\\

\noindent (4) The average star formation rate in the star-forming ETGs in our sample estimated using the radio continuum is $\sim6$ M$_{\odot}$~yr$^{-1}$. This value is consistent with those estimated using FIR and H$\alpha$ emission. \\

\noindent (5) The average value of the magnetic field strengths in star-forming early-type galaxies is estimated to be 12$^{+11}_{-4}$ $\mu$G in the circum-nuclear region for a range of synchrotron path length as $2 - 6$ kpc, a range of source angular size as $7'' - 13''$, and for an assumed radio spectral index as $\sim0.75$.\\

\noindent (6) These magnetic fields are very likely generated via fast amplification in small-scale turbulent dynamos driven by supernovae explosions in the star-bursting regions. \\

It will be worth to take up follow-up studies on this sample of ETGs. The high resolution ($<1$ kpc-scale) multi-frequency radio imaging will be very valuable as it will provide estimates for the spectral index and resolve emissions on sub-galactic scales. The magnetic fields can then be better constrained in compact star-bursting regions. This unique sample of ETGs with high SFR will be an excellent target for radio polarization measurements to understand further evolution of magnetic field, particularly to assess the existence of large-scale magnetic field structure and their possible origin.

\begin{landscape}

\begin{table}
\caption{The estimated values of radio and infrared luminosities, star-formation rates, $'q'$ parameter, thermal-radio fraction, and magnetic field in the radio-detected sample of blue ETGs.}
\centering
\begin{tabular}{lccccccccccc}
\hline
Gal. & S\'ersic$^{a}$ & log & log  & log  & log  & SFR$_{\rm rad}$ & SFR$_{\rm FIR}$ & SFR$_{\rm H\alpha}^{b}$ & $q$ & $R$$_{\rm th}$ & $B$\\
ID & index & ($\frac{L_{radio}^{\rm FIRST}}{\rm W~Hz^{-1}}$) & ($\frac{L_{radio}^{\rm NVSS}}{\rm W~Hz^{-1}}$) & ($\frac{L_{{60\mu}{\rm m}}}{\rm L_\odot}$) & ($\frac{L_{\mathrm{FIR}}}{\rm L_\odot}$)  &M$_{\odot}$~yr$^{-1}$ &M$_{\odot}$~yr$^{-1}$ & M$_{\odot}$~yr$^{-1}$ & & \% & $\mu$G\\
\hline
\multicolumn{12}{@{}l}{\bf Galaxies with S\'ersic index $\ge2$}\\
146 & 8.0 & $21.34\pm0.07$ & $21.52\pm0.13$ & $9.79\pm0.06$ & $10.04\pm0.05$  & $1.8\pm0.5$  & $3.3\pm0.4$ & 4.8 & $2.53\pm0.14$ & $19.4$  & 9.9 \\
84  & 8.0 & $<21.31$       & $<21.78$ & $9.84\pm0.12$ & $<10.36$   & $<3.6$  & $<6.9$  & 1.7 & ---  & ---  & --- \\
14  & 8.0 & x              & $21.99\pm0.09$ & $10.33\pm0.04$ &  $10.54\pm0.04$  & $5.2\pm1.1$  & $10.5\pm1.0$ & 6.1  & $2.55\pm0.10$ & 13.4 & --- \\
50  & 7.3 & $21.98\pm0.02$ & $21.98\pm0.05$ & $10.16\pm0.04$  & $10.39\pm0.04$  & $5.0\pm0.5$  & $7.4\pm0.7$ & 6.6 & $2.41\pm0.07$ & 11.6  & 13.2 \\
61  & 6.3 & $22.00\pm0.02$ & $22.02\pm0.06$ & $10.18\pm0.04$ & $<10.32$  & $5.6\pm0.7$  & $<6.3$  & 21.0 & $<2.30$ & 20.5  & 13.6 \\
195 & 5.4 & $<21.07$       & $<21.54$ & $9.44\pm0.10$ & $9.77\pm0.07$   & $<2.0$  & $1.8\pm0.3$ & 1.2 & $>2.23$  & ---  & --- \\
206 & 5.3 & $<21.46$       & $<21.94$ & $9.78\pm0.17$ &$<10.17$   & $<5.1$  & $<4.5$  & 3.6 & ---  & ---  & --- \\
8   & 4.9 & x              & $21.97\pm0.08$ & $9.84\pm0.18$ &  $<10.39$  & $5.0\pm0.9$  & $<7.4$ & 7.0  & $<2.42$  & 11.8  & --- \\
149 & 4.8 & $22.05\pm0.02$ & $22.11\pm0.04$ & x & x  & $6.8\pm1.2$  & --- & 6.3 & ---  & 7.8  & 15.2 \\
190 & 4.5 & $21.84\pm0.04$ & $22.01\pm0.07$ & $10.11\pm0.05$ & $10.32\pm0.04$  & $5.4\pm0.9$  & $6.3\pm0.6$ & 5.2 & $2.31\pm0.08$ & 8.9  & 11.7 \\
124 & 4.3 & $21.68\pm0.02$ & $21.77\pm0.04$ & $9.86\pm0.03$ & $10.05\pm0.04$   & $3.1\pm0.3$  & $3.4\pm0.3$  & 5.2 & $2.27\pm0.06$ & 11.5  & 14.8 \\
129 & 4.2 & $21.77\pm0.04$ & $22.13\pm0.05$ & $10.26\pm0.03$ & $10.43\pm0.03$  & $7.2\pm0.8$  & $8.2\pm0.6$ & 13.0 & $2.30\pm0.06$ & 12.2  & 12.4 \\
4   & 4.2 & $21.42\pm0.08$ & $<21.69$ & $9.70\pm0.17$ &  $<10.26$  & $>1.4^{c}$  & $<5.5$ & 2.5  & ---  & ---  & 9.4 \\
215 & 4.1 & $21.90\pm0.02$ & $21.88\pm0.07$ & $10.19\pm0.05$ & $10.57\pm0.05$  & $4.0\pm0.7$  & $11.3\pm1.3$ & 7.6 & $2.69\pm0.09$ & 19.6 & 13.5 \\
5   & 3.7 & $<21.29$       & $<21.77$ & $9.64\pm0.16$ & $9.96\pm0.12$   & $<3.5$  & $2.8\pm1.5$ & 3.5 & $>2.2$  & ---  & --- \\
121 & 3.7 & $22.01\pm0.02$ & $22.07\pm0.07$ & $10.01\pm0.07$ & $10.34\pm0.06$  & $6.2\pm1.0$  & $6.6\pm0.9$ & 5.5  & $2.28\pm0.09$ & 8.2  & 13.5 \\
172 & 3.6 & $22.07\pm0.02$ & $21.97\pm0.09$ & $9.91\pm0.13$ & $10.41\pm0.07$   & $5.0\pm1.0$  & $7.8\pm1.3$ & 8.1 & $2.44\pm0.11$ & 13.5  & 14.2 \\
2   & 3.2 & $21.41\pm0.09$ & $<21.77$ & $9.88\pm0.15$ &  $<10.34$  & $>1.3^{c}$  & $<6.6$ & 4.5 & ---  & ---  & 8.9 \\
192 & 2.9 & $<21.34$       & $<21.82$ & $9.68\pm0.16$ & $<10.08$   & $<3.9$  & $<3.6$ & 1.7 & ---  & ---  & --- \\
68  & 2.8 & $21.64\pm0.05$ & $<21.81$ & $<9.6$ & $<10.18$  & $>2.3^{c}$  & $<4.6$  & 3.2 & ---  & ---  & 10.1 \\
177 & 2.8 & $<21.24$       & $<21.71$ & $9.96\pm0.11$ &  $<10.23$   & $<3.0$  & $<5.1$ & 1.5 & ---  & ---  & --- \\
102 & 2.8 & $<21.35$       & $<21.83$ & $9.84\pm0.08$ & $10.16\pm0.06$   & $<4.0$  & $4.4\pm0.6$ & 4.2 & $>2.33$  & ---  & --- \\
56  & 2.7 &     x          & $22.28\pm0.07$ & $10.06\pm0.09$ &  $10.44\pm0.08$  & $10.1\pm1.6$  & $8.3\pm1.6$ & 3.2  & $2.16\pm0.11$ & 4.8  & --- \\
151 & 2.7 & $<21.38$       & $<21.86$ & $9.79\pm0.09$ & $<10.15$   & $<4.3$  & $<4.3$  & 3.7 & ---  & ---  & --- \\
86  & 2.6 & $22.14\pm0.02$ & $22.39\pm0.04$ & $10.14\pm0.07$ & $10.59\pm0.05$  & $13.0\pm1.2$  & $11.8\pm1.4$ & 10.0 & $2.20\pm0.06$ & 7.0  & 12.9 \\
182 & 2.5 & $21.45\pm0.09$ & $22.11\pm0.08$ & $10.12\pm0.04$ & $10.39\pm0.04$  & $6.8\pm1.2$  & $7.4\pm0.7$ & 6.1 & $2.28\pm0.09$ & 8.3  & 8.9 \\
105 & 2.4 & $22.00\pm0.03$ & $22.06\pm0.09$ & $9.93\pm0.07$ & $<10.68$   & $6.1\pm1.2$  & $<14.6$ & 4.0  & $<2.62$ & 5.9  & 12.1 \\
66  & 2.3 & $21.43\pm0.15$ & $<21.91$ & $9.98\pm0.06$ &  $10.23\pm0.07$   & $>1.4^{c}$  & $5.1\pm0.9$ & 6.5 & $>2.32$ & ---  & 8.2 \\
77  & 2.3 & $<21.35$       & $<21.83$ & $9.63\pm0.19$ &  $10.20\pm0.06$   & $<4.0$  & $4.8\pm0.7$ & 0.5 & $>2.37$  & ---  & ---\\
147 & 2.2 & $<21.47$       & $<21.94$ & $10.16\pm0.06$ & $10.57\pm0.04$  & $<5.1$  & $11.3\pm1.1$ & 6.7 & $>2.63$  & ---  & ---\\
175 & 2.1 & $<21.39$       & $<21.87$ & $9.62\pm0.09$ & $10.03\pm0.09$   & $<4.4$  & $3.2\pm0.7$ & 0.6 & $>2.16$  & ---  & ---\\
103 & 2.0 & $21.53\pm0.05$ & $21.88\pm0.07$ & $9.70\pm0.06$ & $10.04\pm0.04$   & $4.0\pm0.7$  & $3.3\pm0.3$  & 3.6  & $2.16\pm0.08$ & 7.3  & 10.7 \\
\hline
\hline
\multicolumn{12}{@{}l}{\bf Galaxies with S\'ersic index $<2$}\\
180 & 1.8 & $22.27\pm0.02$ & $22.30\pm0.04$ & $10.61\pm0.02$ & $10.82\pm0.02$  & $10.6\pm1.0$  & $20.0\pm0.9$ & 18.0 & $2.52\pm0.05$ & 16.5 & 16.3 \\
58  & 1.6 & $21.94\pm0.02$ & $22.09\pm0.04$ & $10.14\pm0.03$ & $10.37\pm0.04$  & $6.6\pm0.6$  & $7.1\pm0.7$  & 5.7 & $2.28\pm0.06$ & 8.3 & 15.4 \\
52  & 1.6 & $21.53\pm0.07$ & $22.07\pm0.07$ & $10.12\pm0.02$ & $10.44\pm0.02$  & $6.2\pm1.0$  & $8.3\pm0.4$  & 3.9 & $2.37\pm0.07$ & 8.5 & 9.4 \\
213 & 1.3 & $21.59\pm0.08$ & $<21.90$ & $10.05\pm0.07$ & $10.30\pm0.08$  & $>2.1^{c}$  & $6.0\pm1.1$ & 7.3 & $>2.40$ & --- & 9.2 \\
148 & 1.3 & $<21.44$ & $<21.92$ & $9.71\pm0.19$ &$<10.15$   & $<4.9$  & $<4.3$ & 1.0 & ---  & ---  & ---\\
7   & 1.0 & x & $22.17\pm0.06$ & $10.48\pm0.04$ & $10.66\pm0.04$  & $7.8\pm1.1$  & $13.9\pm1.3$ & 12.0 & $2.49\pm0.07$ & 13.8 & ---\\
\hline
\multicolumn{12}{@{}l}{Notes: $^{a}$S\'ersic index values are taken from George (2017). $^{b}$The SFR based on H$\alpha$ emission within $3''$ SDSS fibre is directly taken from S09.}\\
\multicolumn{12}{@{}l}{$^{c}$SFRs estimated using FIRST radio flux and are considered as lower limits.}\\
\multicolumn{12}{@{}l}{A 'x' value means no data was available for the given location.}\\

\label{table:tab2}
\end{tabular}
\end{table}

\end{landscape}

\section*{Acknowledgements}
We thank the referee, Aritra Basu for making insightful comments which greatly improved the paper. We acknowledge some preparatory work carried out by P. Aromal and Suvendu Rakshit. The NVSS and FIRST sky survey were conducted by the National Radio Astronomical Observatory (NRAO) using the Very Large Array (VLA). The NRAO is a facility of the National Science Foundation operated under cooperative agreement by Associated Universities, Inc. The NASA/ IPAC Infrared Science Archive, which is operated by the Jet Propulsion Laboratory, California Institute of Technology, under contract with the National Aeronautics and Space Administration. This research has made use of the SAO/NASA Astrophysics Data System (ADS) operated by the Smithsonian Astrophysical Observatory (SAO) under a NASA grant.





\begin{thebibliography}{99}
\bibitem[]{}Adelman-McCarthy J.K. et al., 2008, ApJs, 175, 297 
\bibitem[]{}Arshakian T.G., Beck R., Krause M., Sokoloff D., 2009, A\&A, 494, 21
\bibitem[]{}Baldwin J.A., Phillips M.M., Terlevich R., 1981, PASP, 93, 5 
\bibitem[]{}Bally J., Thronson H.A., 1989, AJ, 97, 69 
\bibitem[]{}Balsara D.S., Kim J., Mac Low M., Mathews G.J., 2004, ApJ, 617, 339 
\bibitem[]{}Basu A., Roy S., Mitra, D., 2012, ApJ, 756, 141
\bibitem[]{}Basu A., Wadadekar Y., Beelen A., Singh V., Archana K.N., Sirothia, S., Ishwara-Chandra C.H., 2015, ApJ, 803, 51
\bibitem[]{}Beck A.M., Lesch H., Dolag K., Kotarba H., Geng A., Stasyszyn F.A., 2012, MNRAS, 422, 2152
\bibitem[]{}Beck R., Brandenburg A., Moss D., Shukurov A., Sokoloff D., 1996, ARA\&A, 34, 155 
\bibitem[]{}Beck R., 2016, Astron. Astrophys. Rev., 24, 4 
\bibitem[]{}Beck R., Krause, M., 2005, Astronomische Nachrichten, 326, 414 
\bibitem[]{}Becker R.H., White R.L., Helfand D.J., 1995, ApJ, 450, 559 
\bibitem[]{}Bernl\"ohr K., 1993, A\&A, 268, 25
\bibitem[]{}Bourne N., Dunne L., Ivison R.J., Maddox S.J., Dickinson M., Frayer D.T., 2011, MNRAS, 410, 1155
\bibitem[]{}Buta R., Combes F., 1996, Fundamentals of Cosmic Physics, 17, 95 
\bibitem[]{}Calzetti D., 2013, in Falcon-Barroso J., Knapen J.H., eds, Secular Evolution of Galaxies, Cambridge Univ. Press, Cambridge, p. 419 
\bibitem[]{}Chy\.zy K.T., 2008, A\&A, 482, 755 
\bibitem[]{}Chy\.zy K.T., Buta R.J., 2008, ApJ, 677, L17
\bibitem[]{}Cappellari M. et al., 2011, MNRAS, 413, 813 
\bibitem[]{}Cappellari M. et al., 2013, MNRAS, 432, 1709  
\bibitem[]{}Colina L., P\'erez-Olea, D.E., 1995, MNRAS, 277, 845 
\bibitem[]{}Combes F., Young L.M., Bureau M., 2007, MNRAS, 377, 1795 
\bibitem[]{}Comer\'on S. et al., 2014, A\&A, 562, A121 
\bibitem[]{}Crocker A.F., Bureau M., Young L.M., Combes F., 2011, MNRAS, 410, 1197 
\bibitem[]{}Condon J.J., 1992, ARA\&A, 30, 575 
\bibitem[]{}Condon J.J., Anderson M.L., Helou G., 1991, ApJ, 376, 95 
\bibitem[]{}Condon J.J., Cotton W.D., Greisen E.W., Yin Q.F., Perley R.A., Taylor G.B., Broderick J.J., 1998, AJ, 115, 1693 
\bibitem[]{}Condon J.J., Cotton W.D., Broderick J.J., 2002, ApJ, 124, 675 
\bibitem[]{}Davis T.A. et al., 2014, MNRAS, 444, 3427
\bibitem[]{}Dopita M.A., Pereira M., Kewley L.J., Capaccioli M., 2002, ApJs, 143, 47 
\bibitem[]{}Durrer R., Neronov A., 2013, Astron. Astroph. Rev, 21, 62 
\bibitem[]{}Fitt A.J., Alexander P., 1993, MNRAS, 261, 445 
\bibitem[]{}George K., 2017, A\&A, 598, A45  
\bibitem[]{}Govoni F., Feretti L., 2004, Intl. J. Mod. Phys. D., 13, 1549
\bibitem[]{}Gunn, J.E. et al., 2006, AJ, 131, 2332
\bibitem[]{}Hanasz M., Woltanski D., Kowalik K., 2009, ApJ, 706, L155
\bibitem[]{}Hanayama H., Takahashi K., Kotake K., Oguri M., Ichiki K., Ohno H., 2005, ApJ, 633, 941   
\bibitem[]{}Harwit M., Pacini F., 1975, ApJ, 200, L127 
\bibitem[]{}Helou G., 1986, ApJ, 311, 33 
\bibitem[]{}Helou G., Bicay M.D., 1993, ApJ, 415, 93 
\bibitem[]{}Helou G., Kahn I.R., Malek L., Boehmer L., 1988, ApJS, 68, 151 
\bibitem[]{}Hill T.L., Heisler C.A., Norris R.P., Reynolds J.E., Hunstead R.W., 2001, AJ, 121, 128 
\bibitem[]{}Inoue A.K., Hirashita H., Kamaya H., 2000, PASJ, 52, 539
\bibitem[]{}Ivison R.J. et al., 2010, A\&A, 518, L31 
\bibitem[]{}Jaiswal S., Omar A., 2016, MNRAS, 462, 92 
\bibitem[]{}Jarvis M.J. et al., 2010, MNRAS, 409, 92
\bibitem[]{}Jura M., 1986, ApJ, 306, 483 
\bibitem[]{}Kauffmann G. et al., 2003, MNRAS, 346, 1055 
\bibitem[]{}Kennicutt, R.C., 1998, ApJ, 498, 541 
\bibitem[]{}Kormendy J., 1977, ApJ, 218, 333
\bibitem[]{}Kostiuk I.P., Silchenko O.K., 2015, Astrophys. Bull., 70, 280 
\bibitem[]{}Krause M., Wielebinski R., Dumke M., 2006, Astronomische Nachrichten, 327, 499 
\bibitem[]{}Kroupa P., 2002, Science, 295, 82
\bibitem[]{}La Barbera F., Ferreras I., Vazdekis A., 2015, MNRAS, 449, L137 
\bibitem[]{}Lacki B.C., Thompson T.A., Quataert E., 2010, ApJ, 717, 1 
\bibitem[]{}Lazar M., Schlickeiser R., Wielebinski R., Poedts S., 2009, ApJ, 693, 1133 
\bibitem[]{}Lisenfeld U., V\"olk H.J., Xu C., 1996, A\&A, 306, 677 
\bibitem[]{}Lisenfeld U. et al., 2007, A\&A, 462, 507 
\bibitem[]{}Longair M.S., 1994, In High Energy Astrophysics, 2nd edition, Cambridge Univ. Press, vol. 2
\bibitem[]{}Lucero D.M., Young, L.M., 2007, AJ, 134, 2148 
\bibitem[]{}Mihos J.C., Hernquist L., 1994, ApJ, 431, L9
\bibitem[]{}Miller N.A., Owen F.N., 2001, AJ, 121 1903 
\bibitem[]{}Mori\'c I., Smol\v{c}i\'c V., Kimball A., Riechers D.A., Ivezi\'c, \v{Z}., 2010, ApJ, 724, 779 
\bibitem[]{}Moss D., Mikhailov E., Silchenko O., Sokoloff D., Horellou C., Beck R., 2016, A\&A, 592, A44 
\bibitem[]{}Murgia M., Helfer T.T., Ekers R., Blitz L., Moscadelli L., Wong, T., Paladino R., 2005, A\&A, 437, 389 
\bibitem[]{}Neugebauer G. et al., 1984, ApJ, 278, L1 
\bibitem[]{}Niklas S., Beck R., 1997, A\&A, 320, 54 
\bibitem[]{}Nyland K. et al., 2017, MNRAS, 464, 1029 
\bibitem[]{}Omar A., Dwarakanth K.S., 2005, JAA, 26, 89
\bibitem[]{}Paswan A., Omar A., 2016, MNRAS, 459, 233 
\bibitem[]{}Planck Collaboration, 2016, A\&A, 594, A13 
\bibitem[]{}Pogge R.W., Eskridge P.B., 1993, AJ, 106, 1405 
\bibitem[]{}Reddy N.A., Yun, M.S., 2004, ApJ, 600, 695 
\bibitem[]{}Rees M., 2005, Lecture Notes in Physics, 664, 1 
\bibitem[]{}Roychowdhury S., Chengalur J.N., 2012, MNRAS, 423, L127 
\bibitem[]{}Ryle M., Hewish A., 1960, MNRAS, 120, 220 
\bibitem[]{}Salim S., Fang J.J., Rich R.M., Faber S.M., Thilker D.A., 2012, ApJ, 755,105 
\bibitem[]{}Sargent M.T. et al., 2010, ApJs,186, 341 
\bibitem[]{}Schawinski K., Khochfar S., Kaviraj S., 2006, Nature, 442, 888 
\bibitem[]{}Schawinski K., Thomas D., Sarzi M., Maraston C., Kaviraj S., Joo S., Yi S.K., Silk J., 2007, MNRAS, 382, 1415
\bibitem[]{}Schawinski K. et al., 2009, MNRAS, 396, 818 
\bibitem[]{}Schleicher D.R.G., Beck R., 2013, A\&A, 556, A142 
\bibitem[]{}Schlickeiser R., Felten T., 2013, ApJ, 778, 39 
\bibitem[]{}Schober J., Schleicher D.R.G., Federrath C., Klessen R.S., Banerjee R., 2012, Phys. Rev. E, 85, 26303
\bibitem[]{}Schober J., Schleicher D.R.G., Klessen R.S., 2016, ApJ, 827, 109
\bibitem[]{}Schommer R.A., Sullivan W.T., 1976, Astrophys. Lett., 17, 191 
\bibitem[]{}Suchkov A., Allen R.J., Heckman T.M., 1993, ApJ, 413, 542
\bibitem[]{}Tabatabaei F.S. et al., 2013, A\&A, 552, A19
\bibitem[]{}Thompson T.A., Quataert E., Waxman E., Murray N., Martin C.L., 2006, ApJ, 645, 186 
\bibitem[]{}van der Kruit P.C., 1971, A\&A, 15, 110 
\bibitem[]{}van der Kruit P.C., 1973, A\&A, 29, 263 
\bibitem[]{}V\"olk H.J., 1989, A\&A, 218, 67 
\bibitem[]{}Walsh D.E.P., Knapp G.R., Wrobel J.M., Kim, D.-W., 1989, ApJ, 337, 209 
\bibitem[]{}White R.L., Becker R.H., Helfand D.J., Gregg M.D., 1997, ApJ, 475, 479 
\bibitem[]{}Wrobel J.M., Heeschen D.S., 1991, AJ, 101, 148 
\bibitem[]{}Xu C., 1990, ApJ, 365, 47 
\bibitem[]{}Yun M.S., Reddy N.A., Condon J.J., 2001, ApJ, 554, 803 
\end{thebibliography}


\appendix

\section{Estimation of luminosities, thermal radio flux, star formation rates, and magnetic field strengths}

A brief overview of the methods used to estimate different parameters presented in this paper is provided here. 

The radio and far-infrared flux densities were converted to luminosities using the relations:

\begin{equation}
{\rm log}~\frac{L_{\rm 1.4GHz}}{\rm W~Hz^{-1}} = 20.08 + 2~{\rm log}~\frac{D}{\rm Mpc} + {\rm log}~ \frac{S_{\rm 1.4GHz}}{\rm Jy}
\label{equ:RADLUM}
\end{equation}

\begin{equation}
{\rm log}~\frac{L_{\rm FIR}}{\rm L_{\odot}} = 5.602 + 2~{\rm log}~\frac{D}{\rm Mpc} +
{\rm log}~\frac{2.58~S_{{60\mu}{\rm m}} + S_{{100\mu}{\rm m}}}{\rm Jy}
\label{equ:FIRLUM}
\end{equation}

\noindent where flux and luminosity are denoted by $S$ and $L$ with a subscript indicating radio or infrared wave-band. $D$ represents the distance to the galaxy. Since the redshifts ($z$) of the galaxies studied here are low ($1 + z \approx$ 1), no $k$-correction was performed. 

Following Helou et al. (1998), the $'q'$ parameter was defined as: 

\begin{equation}
q = {\rm log} \left (\frac{2.58~S_{{60\mu}{\rm m}} + S_{{100\mu}{\rm m}}}{2.98~\rm Jy} \right) - {\rm log} \left(\frac{S_{\rm 1.4GHz}}{\rm Jy} \right)
\label{equ:qpar}
\end{equation}

or equivalently, 

\begin{equation}
q \approx 14.0 + {\rm log}~\frac{L_{\rm FIR}}{\rm L_{\odot}} - {\rm log}~\frac{L_{\rm 1.4GHz}}{\rm W~Hz^{-1}}
\end{equation}

Kennicutt (1998) SFR relations were used to derive ${\rm H\!\alpha}$ flux from the ${\rm H\!\alpha}$ SFR provided in S09. The FIR luminosity was converted to SFR and also used to estimate equivalent ${\rm H\!\alpha}$ flux using the Kennicutt (1998) SFR relations assuming that both FIR and ${\rm H\!\alpha}$ trace same SFR:

\begin{equation}
\frac{\rm SFR}{\rm ~M_\odot~yr^{-1}} = {{L_{\rm FIR} (\rm L_{\odot})} \over {3.3 \times 10^9} } 
\end{equation}

\begin{equation}
\frac{\rm SFR}{\rm ~M_\odot~yr^{-1}} = 9\times 10^8 \left(\frac{D}{\rm Mpc} \right)^2 \left(\frac{S_{\rm H\!\alpha}}{\rm erg~s^{-1}~cm^{2}}\right)
\end{equation}

Subsequently, two values for the thermal-radio component for each galaxy were estimated, one value using the derived $\rm {H\!\alpha}$ flux from the FIR luminosity and other using the directly observed $\rm {H\!\alpha}$ flux from S09 and the relation given by Dopita et al. (2002):

\begin{equation}
\frac{S_{\rm 1.4GHz}^{\rm thermal}}{\rm Jy} = 1.21 \times 10^{9} ~\left(\frac{S_{\rm H\!\alpha}}{\rm erg~s^{-1} cm^{2}}\right)
\end{equation}

\noindent where $S_{\rm H\!\alpha}$ denotes $\rm {H\!\alpha}$ flux and SFR is the star formation rate. The SFR estimates are generally valid for a constant SFR in the past 100 Myr and are sensitive to variations in the initial mass function and SF history. 

The thermal-radio fraction (in per cent) was estimated as:

\begin{equation}
R_{\rm th} = 100~\frac{S_{\rm 1.4GHz}^{\rm thermal}}{S_{\rm 1.4GHz}^{\rm NVSS}}
\end{equation}

The SFR using the total (thermal plus non-thermal) 1.4 GHz radio flux density was estimated using the relation given in Yun et al. (2001):

\begin{equation}
\frac{\rm SFR}{\rm ~M_\odot~yr^{-1}} \approx 5.9 (\pm1.8) \times 10^{-22} \frac{L_{\rm 1.4GHz}}{\rm W~Hz^{-1}}
\end{equation}
 
The magnetic field ($B$) at low redshifts ($1+z\approx1$) can be estimated using the following relation obtained after minimizing the total energy density (CR and magnetic field) in the ISM: 

\begin{equation}
B (\mu {\rm G}) = A_{\alpha,\nu_{min},\nu_{max}}~(1 + K)^{\beta}~\nu^{\alpha\beta}~I^{\beta}~d^{-\beta}
\end{equation}

\noindent where $\alpha$ is radio spectral index, $\nu$ is radio frequency in MHz, $I$ is radio surface flux density in mJy arcsec$^{-2}$ measured at frequency $\nu$ and $d$ is the path length of the synchrotron emitting region in kpc. In classical derivations (see Longair 1994) the index $\beta$ is constant at a value of 2/7. In the modified estimates of Beck \& Krause (2005), the value of $\beta$ is 1/(3+$\alpha$). $K$ is the ratio of the total energy of CR nuclei to that of the synchrotron emitting electrons and positrons in the classical formulae while $K$ is taken as the ratio of number densities of CR protons and electrons per particle energy interval within the energy range traced by the observed synchrotron emission. The $K$ is normally taken as a constant value at 100, consistent with the values measured in the local Galactic CR data near the sun at particle energies of a few GeV and predictions from Fermi shock acceleration and hadronic interaction models (see Beck \& Krause 2005 for details).  The factor $A$($\alpha,\nu_{min},\nu_{max}$) can be estimated from Govoni \& Feretti (2004) and Beck \& Krause (2005) for the classical and modified cases respectively. 

The spectral indices $\alpha$ are not known for the galaxies studied here. We assumed $\alpha$ as $\sim0.75$ consistent with the average values of radio spectral index observed in the star-forming galaxies (Condon 1992). For $K$ = 100 and $\alpha\sim0.75$, we found that the derived field strengths using formulae with classical index 2/7 and modified index 1/(3+$\alpha$) give nearly identical results. At $\alpha\sim0.75$, the uncertainties in the field estimates due to unknown values of path length and radio source size in our sample are much higher than the uncertainty arising from the use of the classical formulae or the modified formulae. Therefore, the field strengths can be estimated using the classical formulae or the modified formulae for $\alpha \approx 0.75$, without introducing a significant error. In the absence of spectral-index measurements, the simplified classical relation to estimate galactic magnetic field in terms of the synchrotron surface flux density measured at 1.4 GHz, and assuming $\alpha\sim0.75$, can be written as: 

\begin{equation}
B (\mu {\rm G}) \approx 60 \left(\frac{I_{\rm {1.4 GHz}}}{\rm {mJy~arcsec^{-2}}}\right)^{2/7} \left(\frac{d}{\mathrm{kpc}}\right)^{-2/7}
\end{equation}

It is to be noted here that at significantly flatter or steeper spectral index than 0.75, the above equation requires significant corrections as pointed out in Beck \& Krause (2005).

\section{Radio images}

The remaining radio-contour images overlaid upon the grey-scale SDSS g-band images are presented here. The images are made following the same contour scheme as in the Fig.~1.  

\begin{figure*}
\centering
\includegraphics[width=7.5cm,trim={2cm 3.2cm 3.8cm 3.3cm},clip]{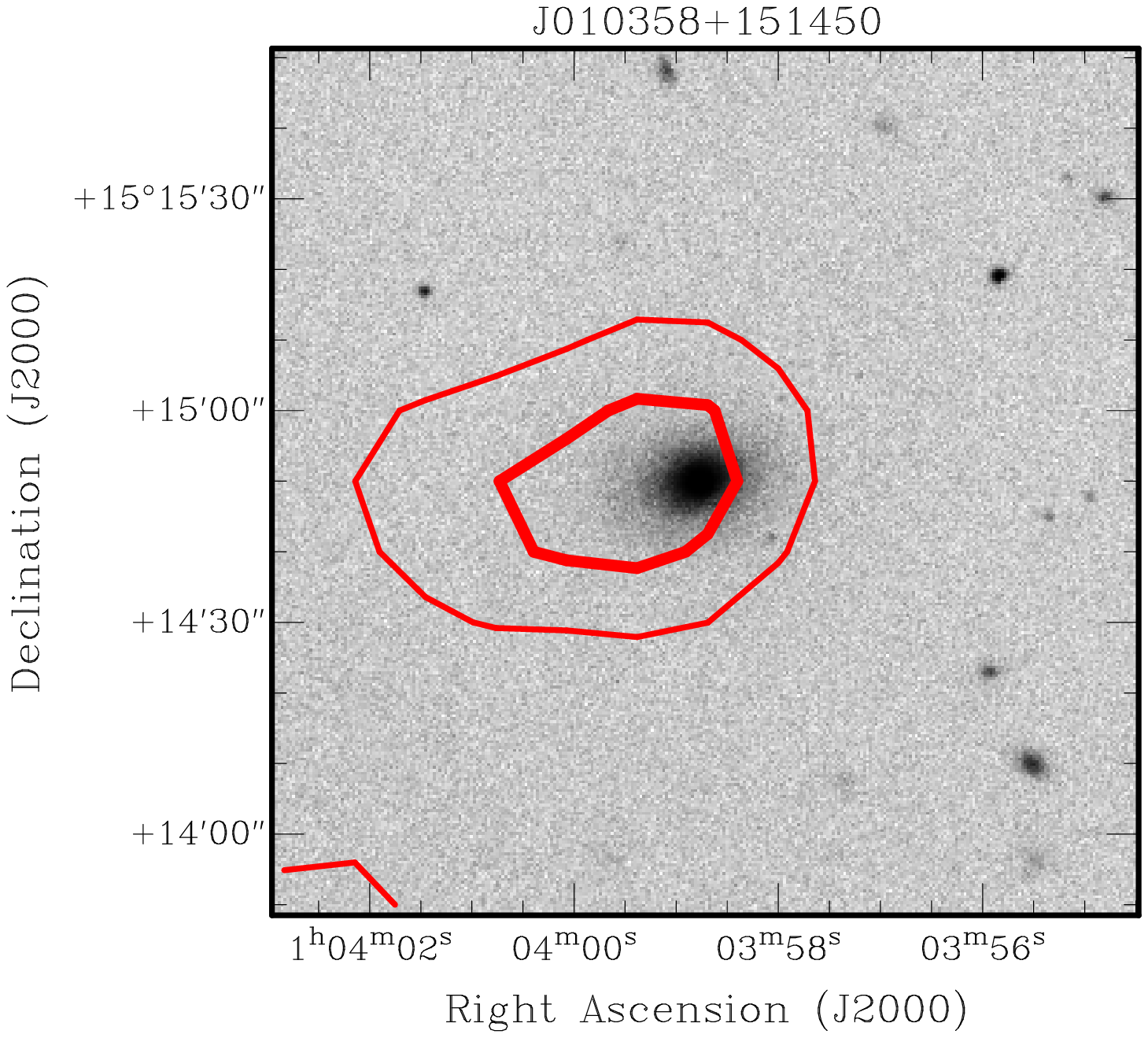}
\includegraphics[width=7.5cm,trim={2cm 3.2cm 3.8cm 3.3cm},clip]{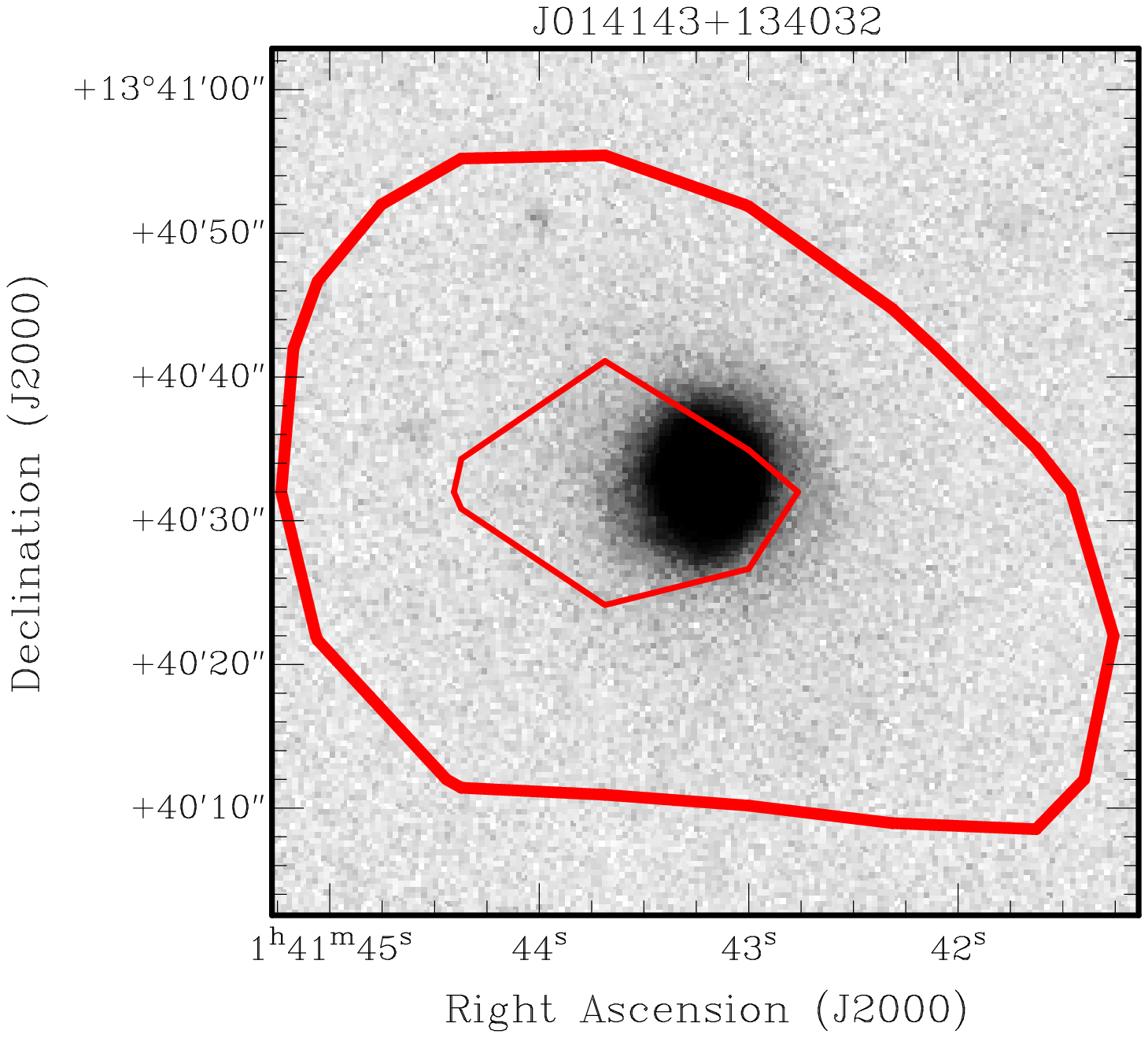}

\vspace{0.15in}
\includegraphics[width=7.5cm,trim={2cm 3.2cm 3.8cm 3.3cm},clip]{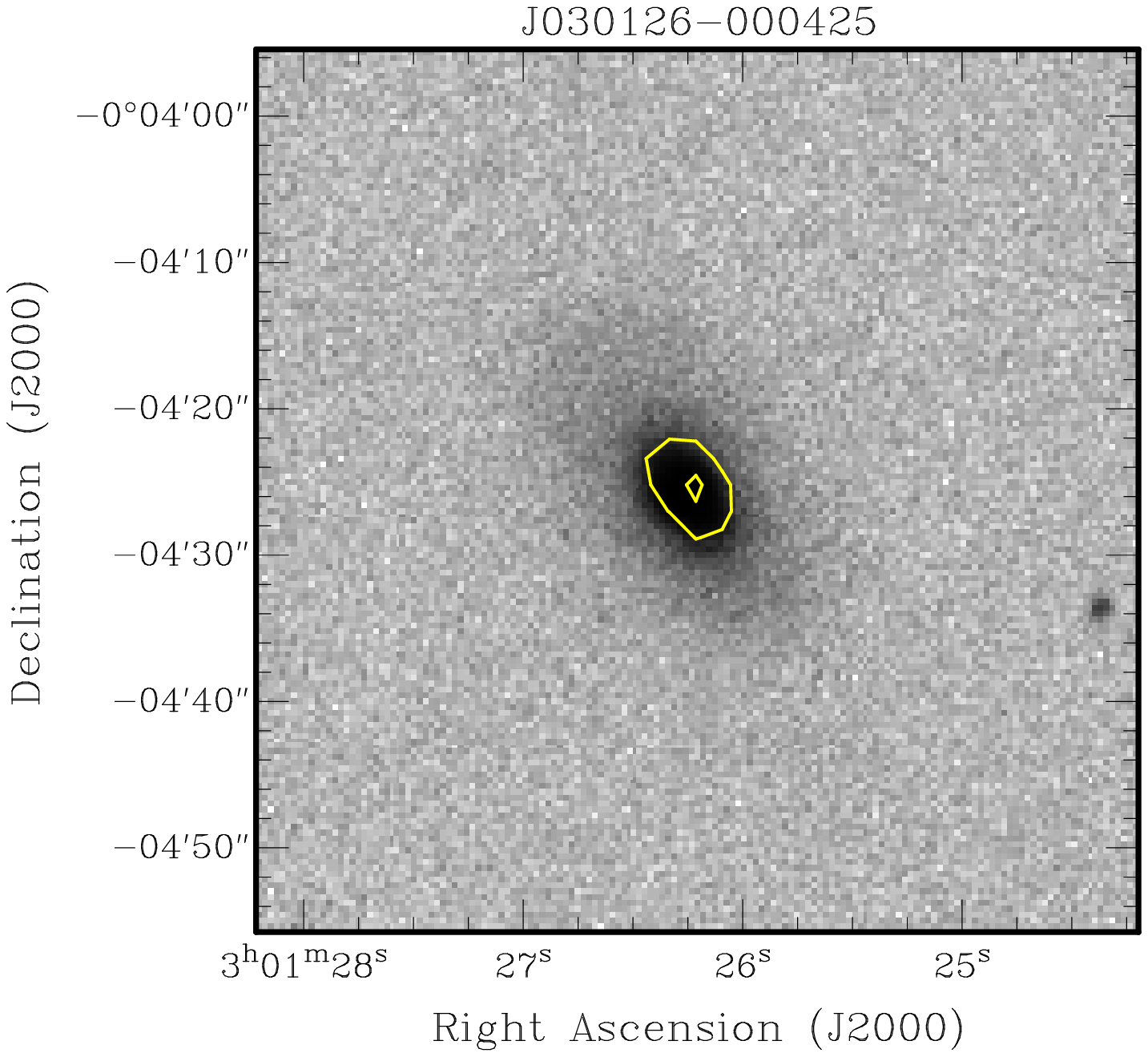}
\includegraphics[width=7.5cm,trim={2cm 3.2cm 3.8cm 3.3cm},clip]{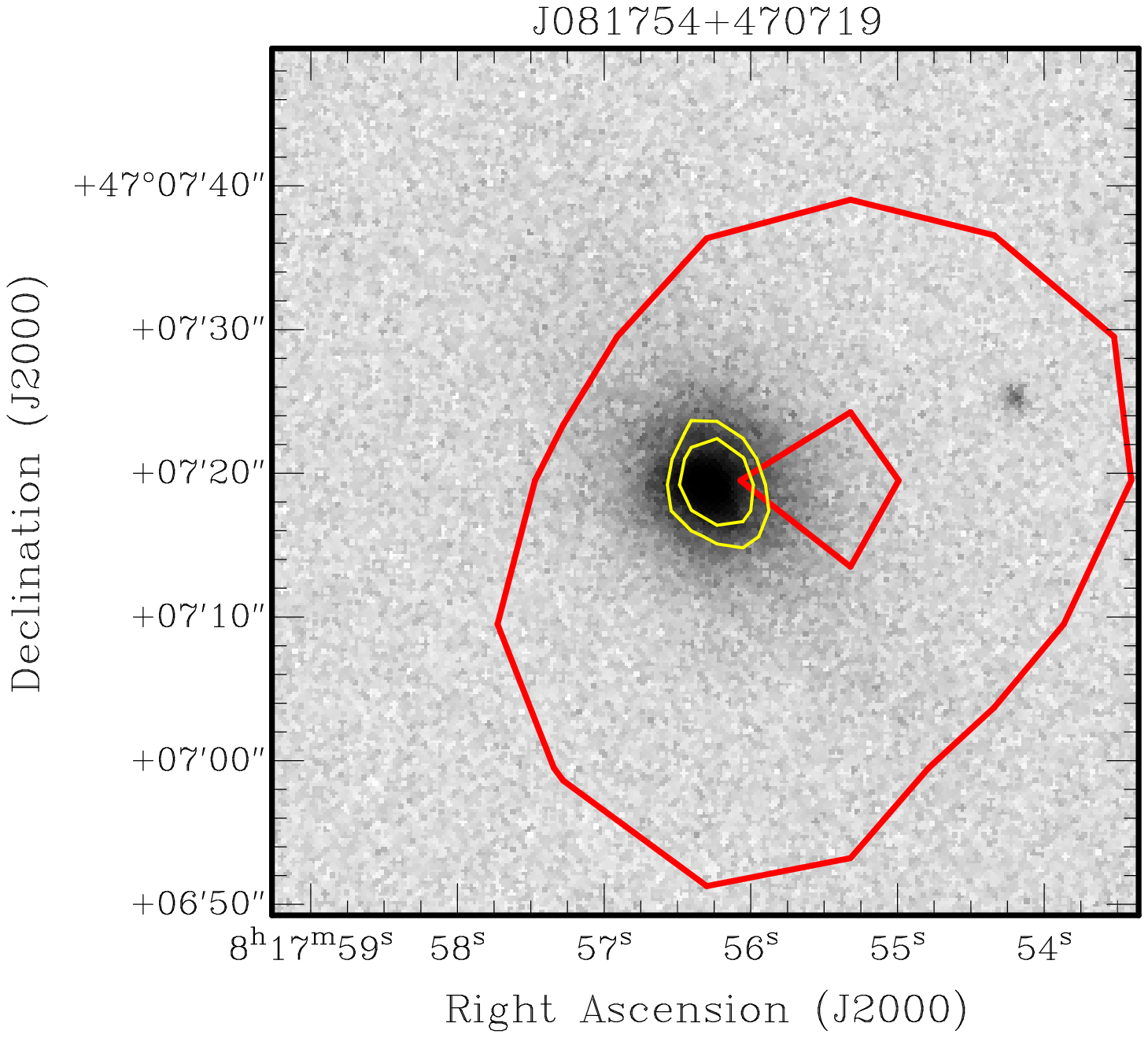}

\vspace{0.15in}
\includegraphics[width=7.5cm,trim={2cm 3.2cm 3.8cm 3.3cm},clip]{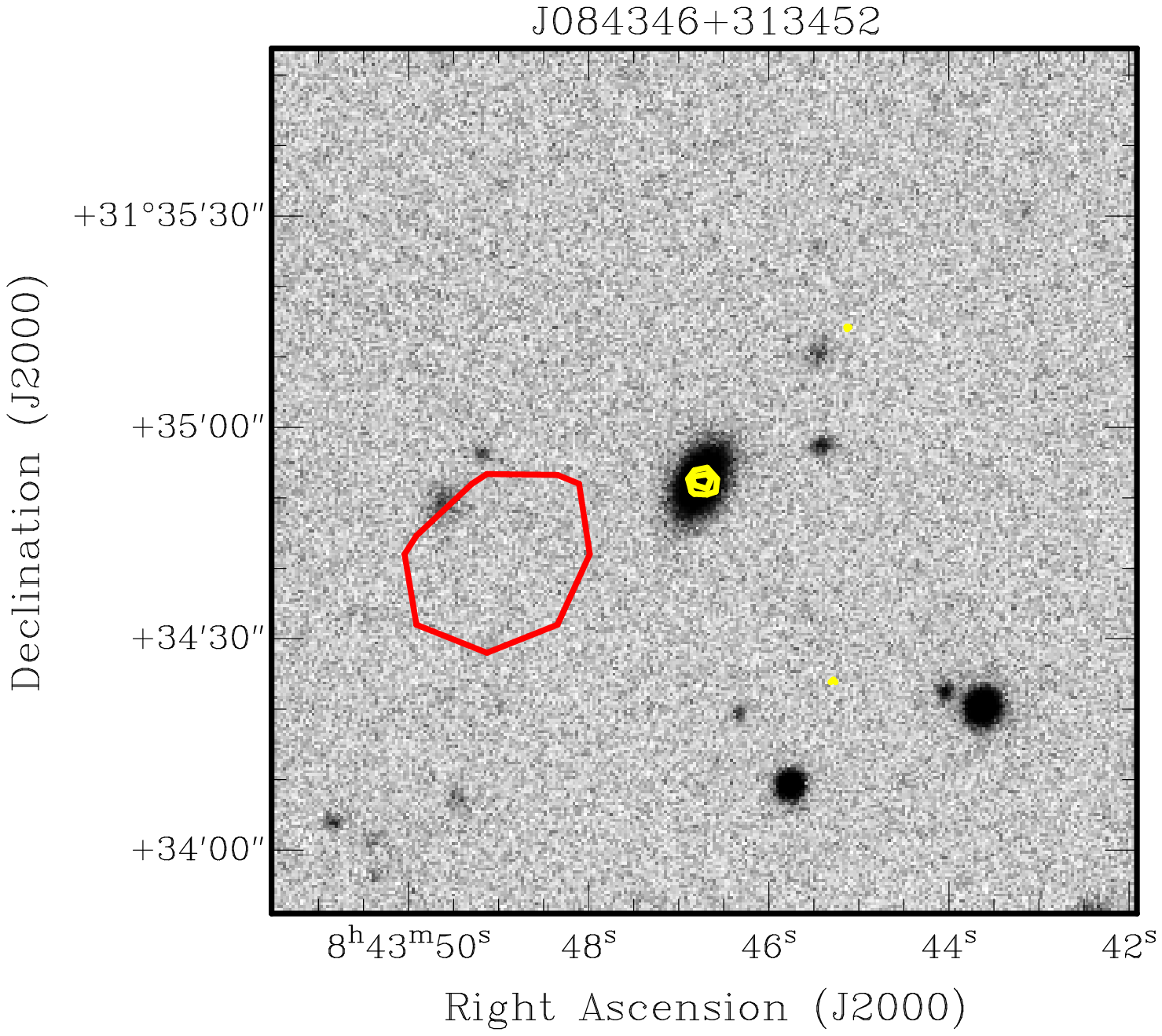}
\includegraphics[width=7.5cm,trim={2cm 3.2cm 3.8cm 3.3cm},clip]{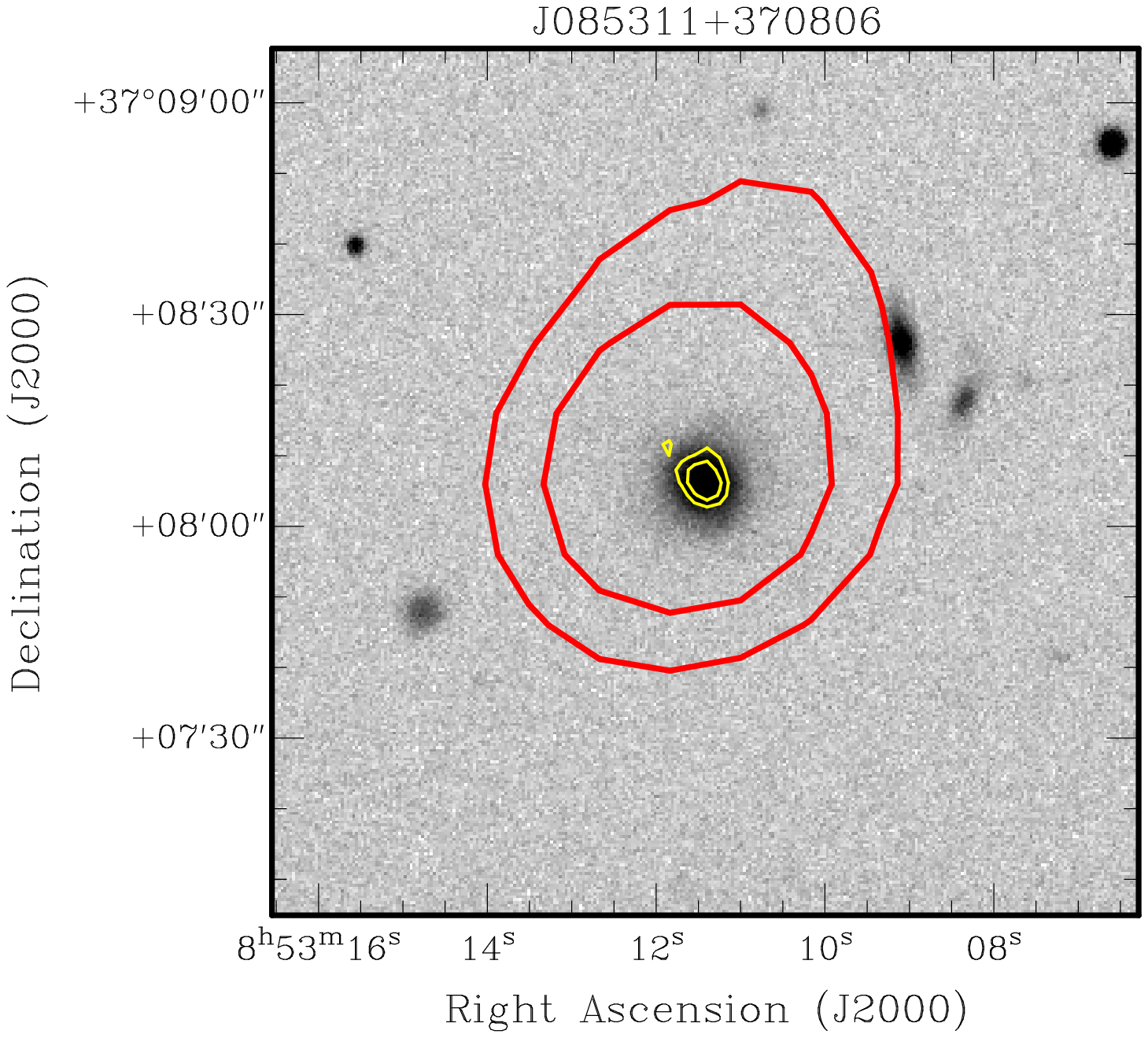}
\end{figure*}

\begin{figure*}
\centering
\includegraphics[width=7.5cm,trim={2cm 3.2cm 3.8cm 3.3cm},clip]{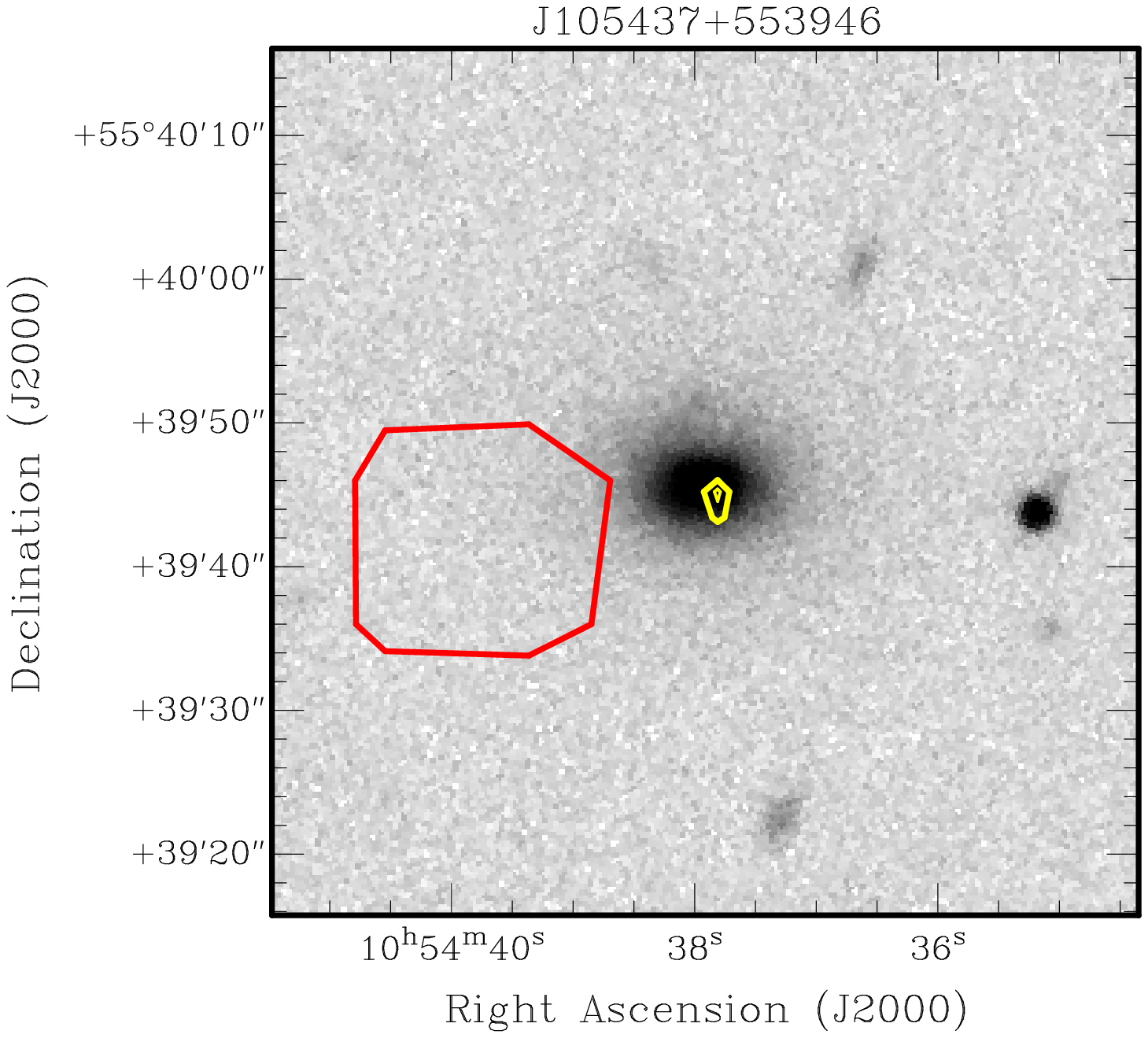}
\includegraphics[width=7.5cm,trim={2cm 3.2cm 3.8cm 3.3cm},clip]{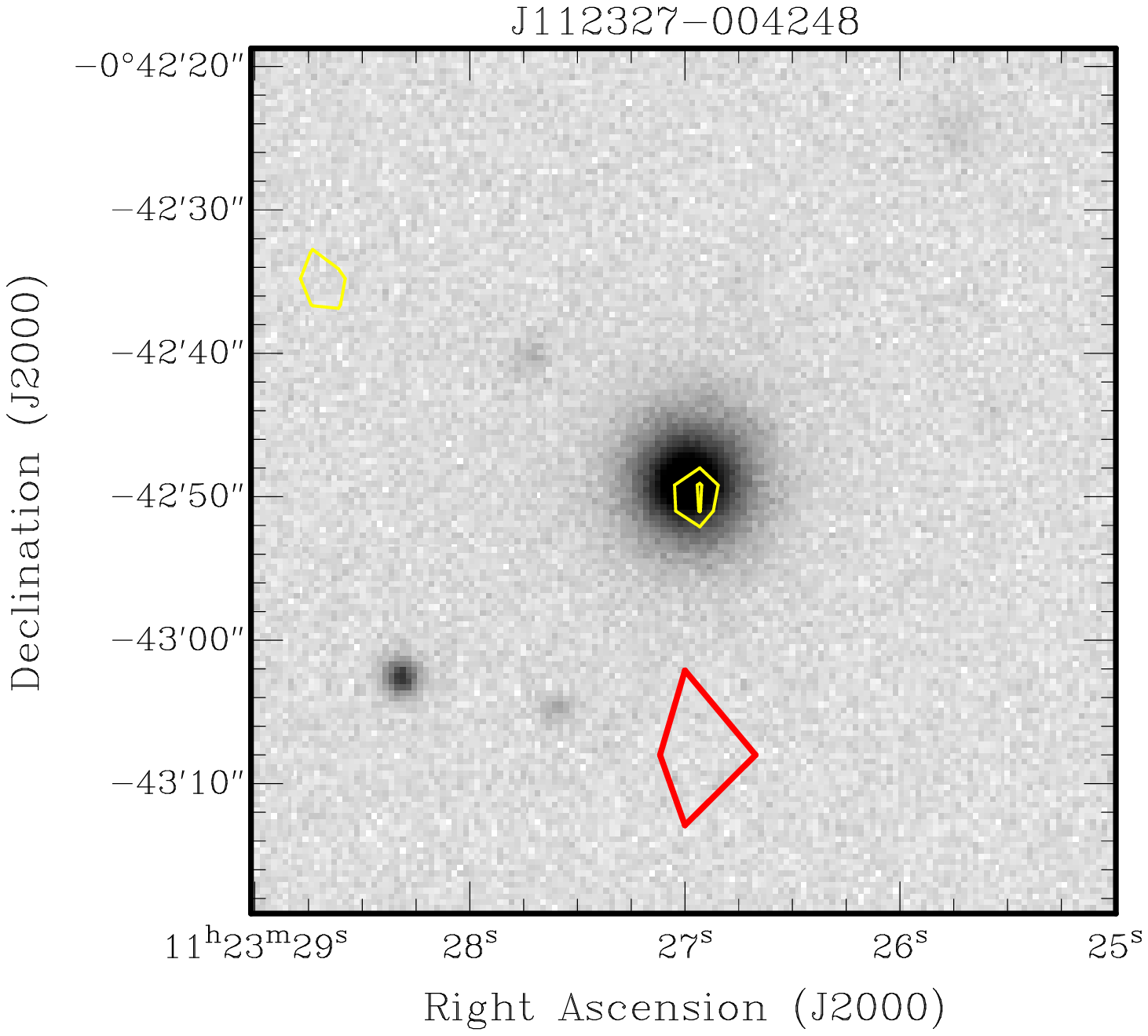}

\vspace{0.15in}
\includegraphics[width=7.5cm,trim={2cm 3.2cm 3.8cm 3.3cm},clip]{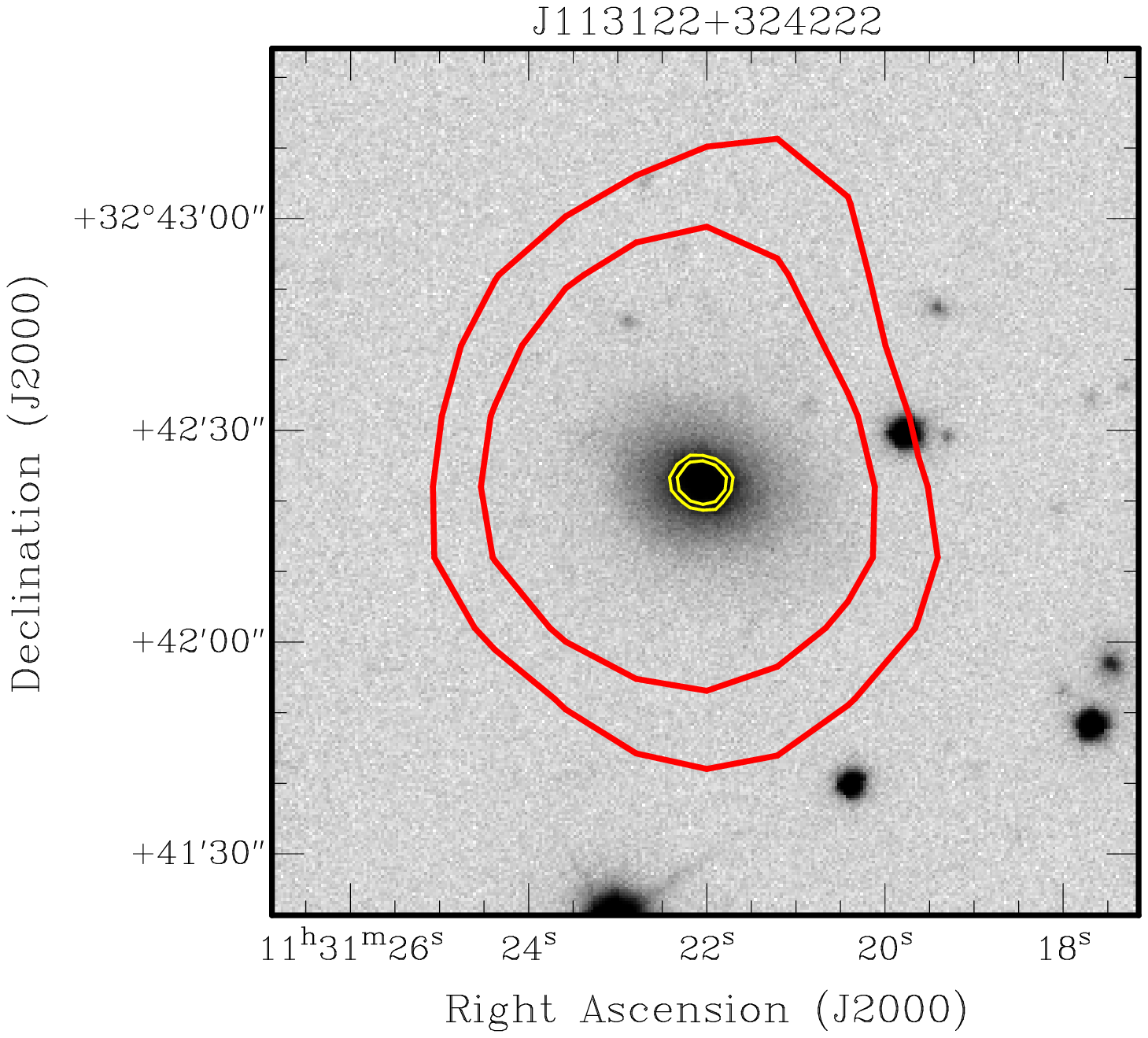}
\includegraphics[width=7.5cm,trim={2cm 3.2cm 3.8cm 3.3cm},clip]{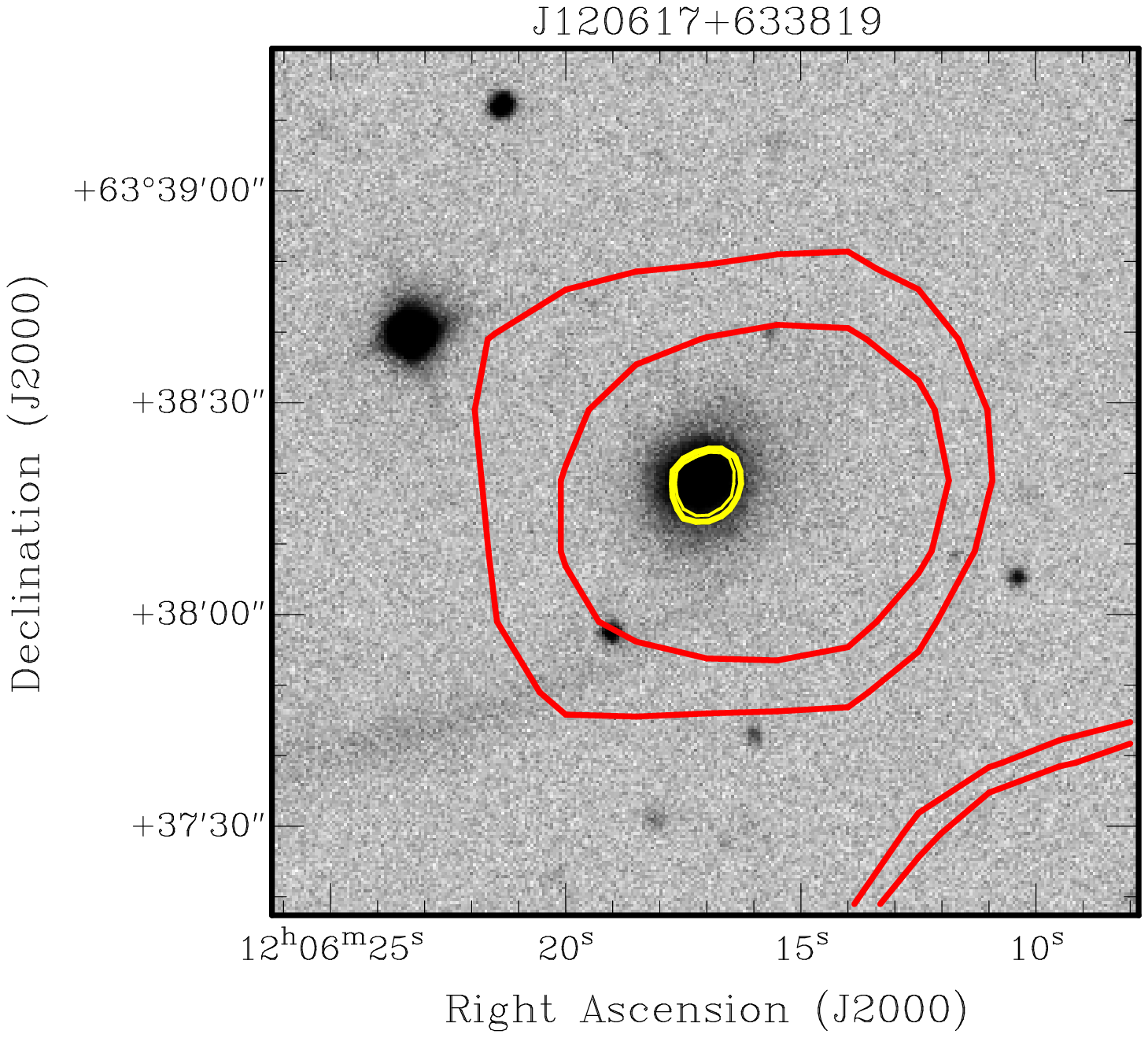}

\vspace{0.15in}
\includegraphics[width=7.5cm,trim={2cm 3.2cm 3.8cm 3.3cm},clip]{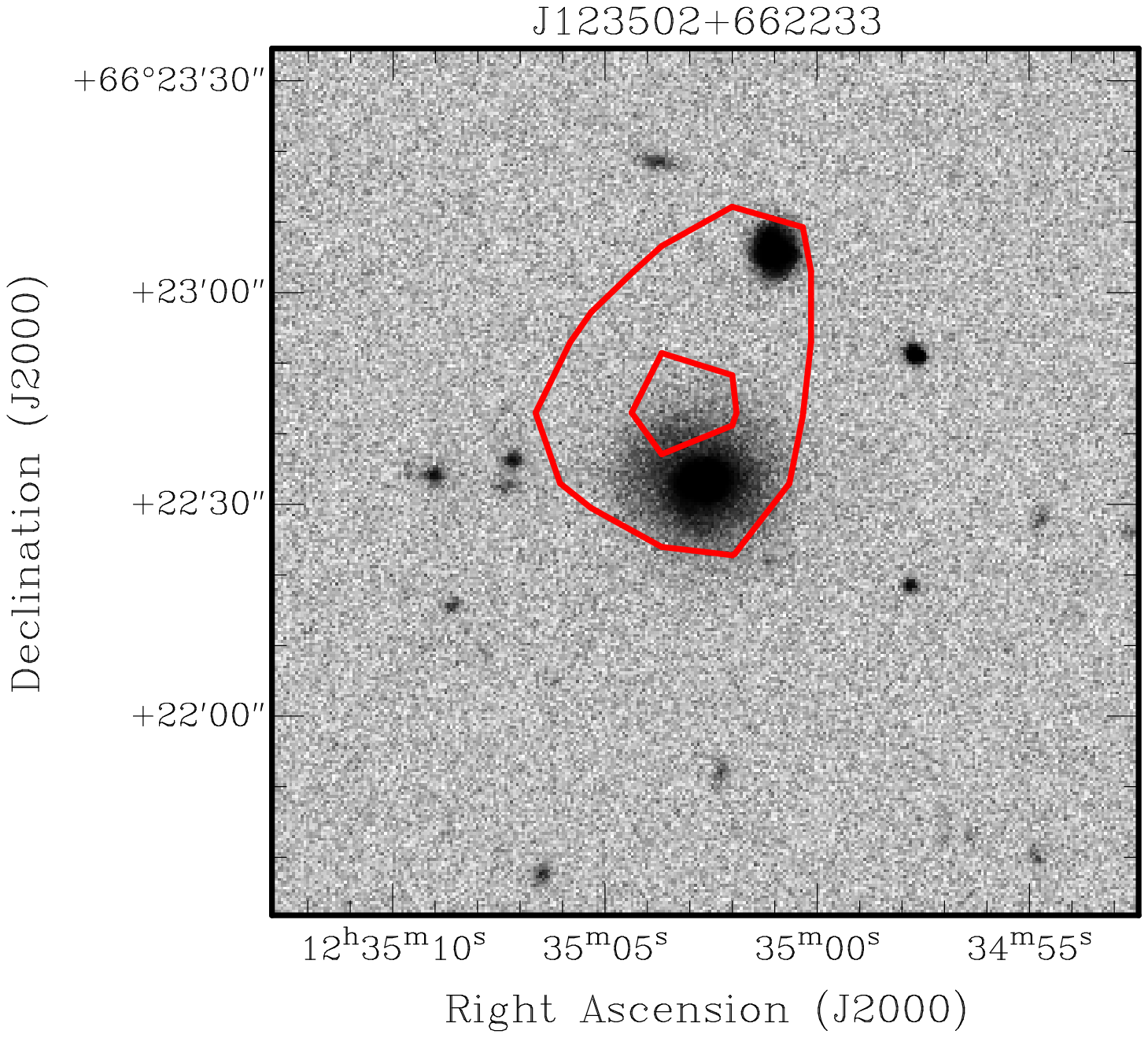}
\includegraphics[width=7.5cm,trim={2cm 3.2cm 3.8cm 3.3cm},clip]{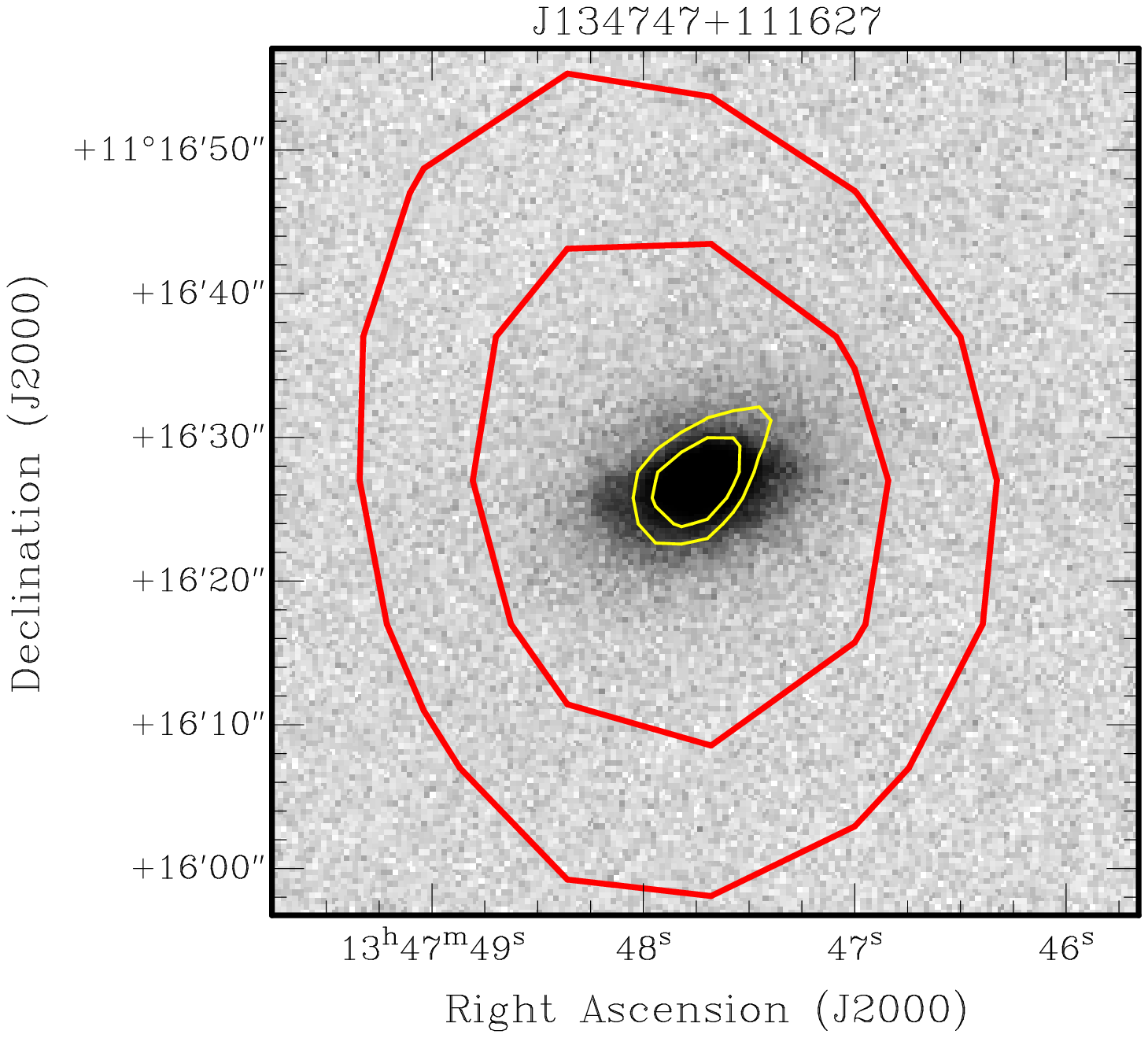}
\end{figure*}

\begin{figure*}
\centering
\includegraphics[width=7.5cm,trim={2cm 3.2cm 3.8cm 3.3cm},clip]{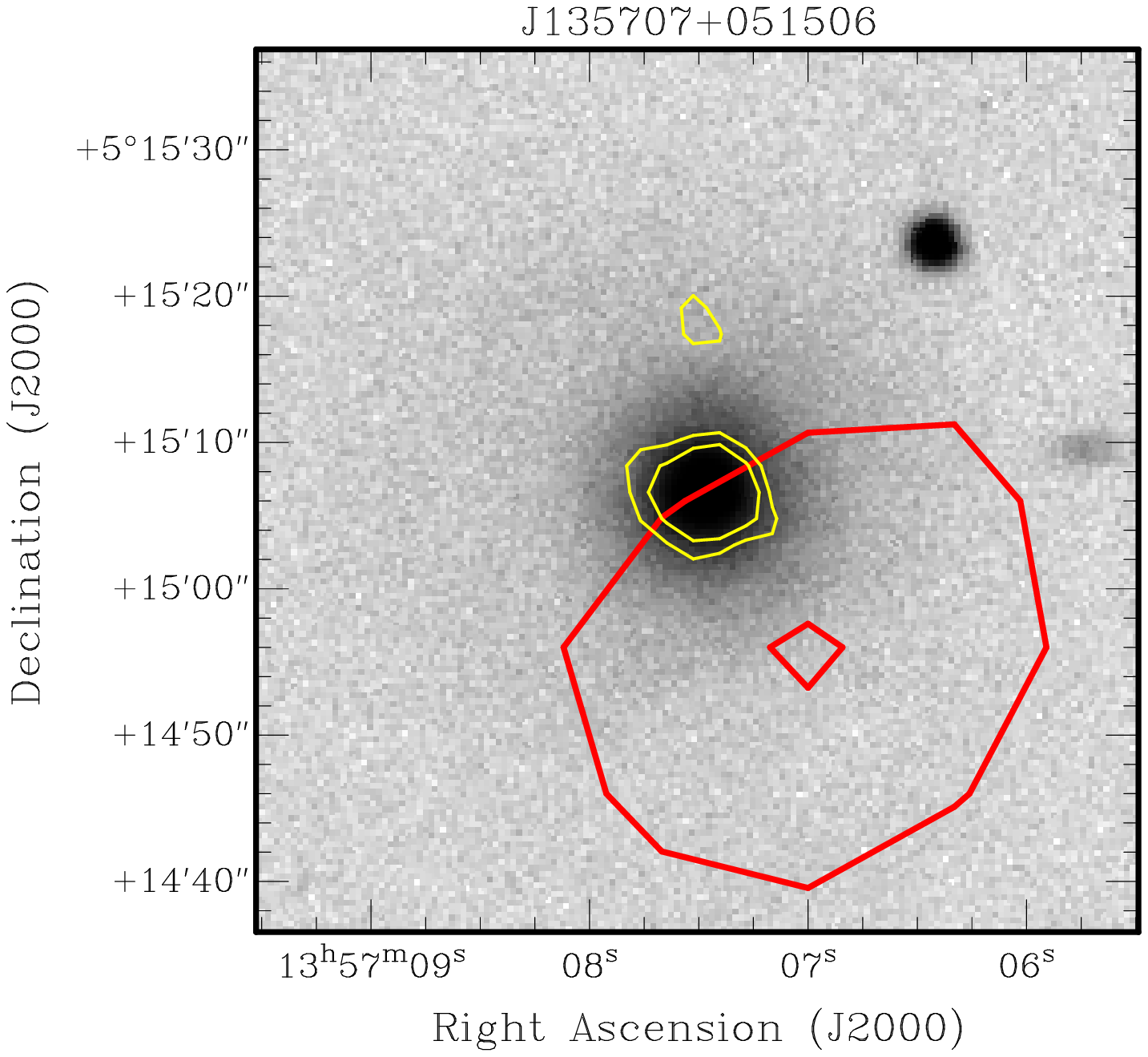}
\includegraphics[width=7.5cm,trim={2cm 3.2cm 3.8cm 3.3cm},clip]{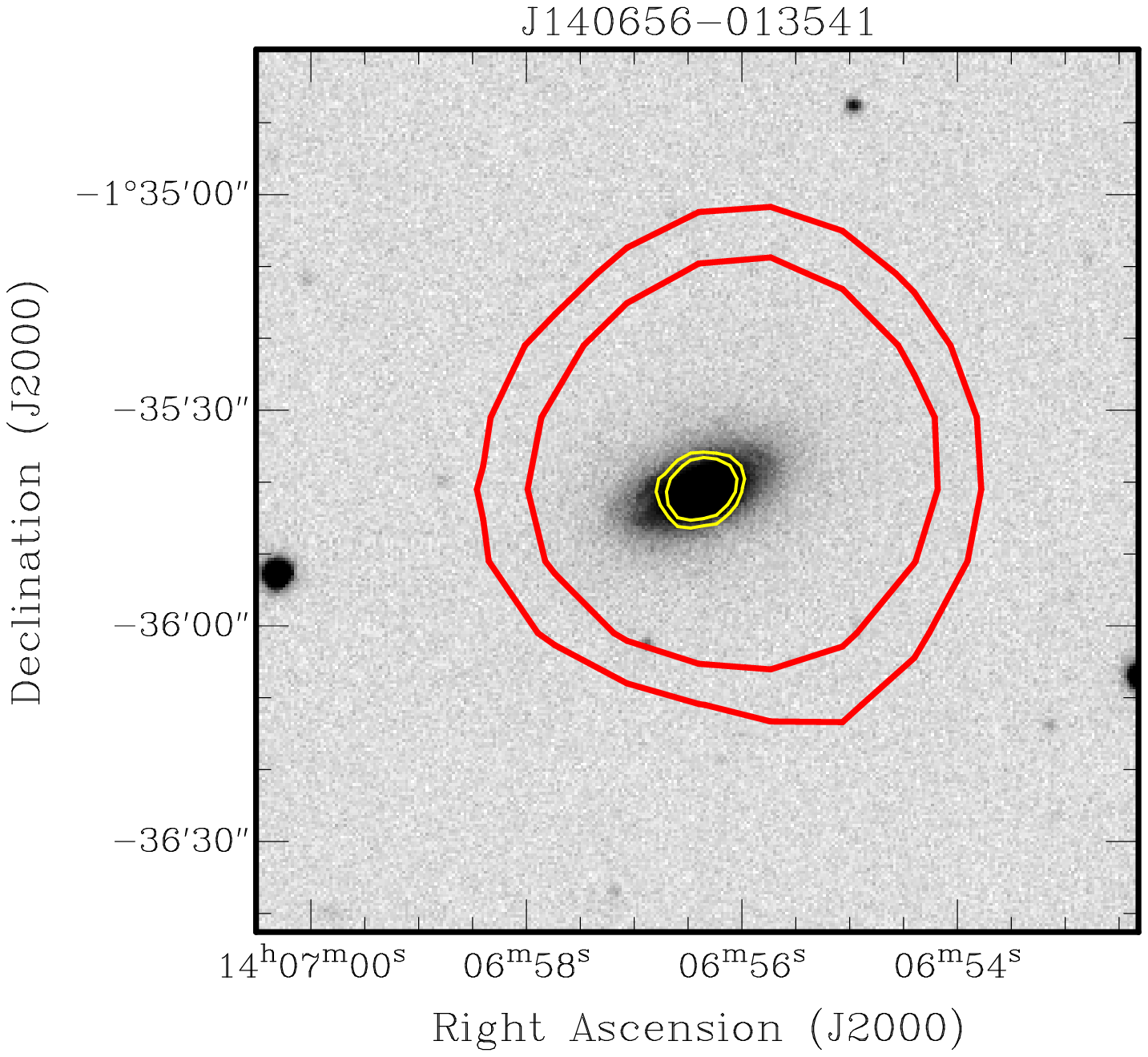}

\vspace{0.15in}
\includegraphics[width=7.5cm,trim={2cm 3.2cm 3.8cm 3.3cm},clip]{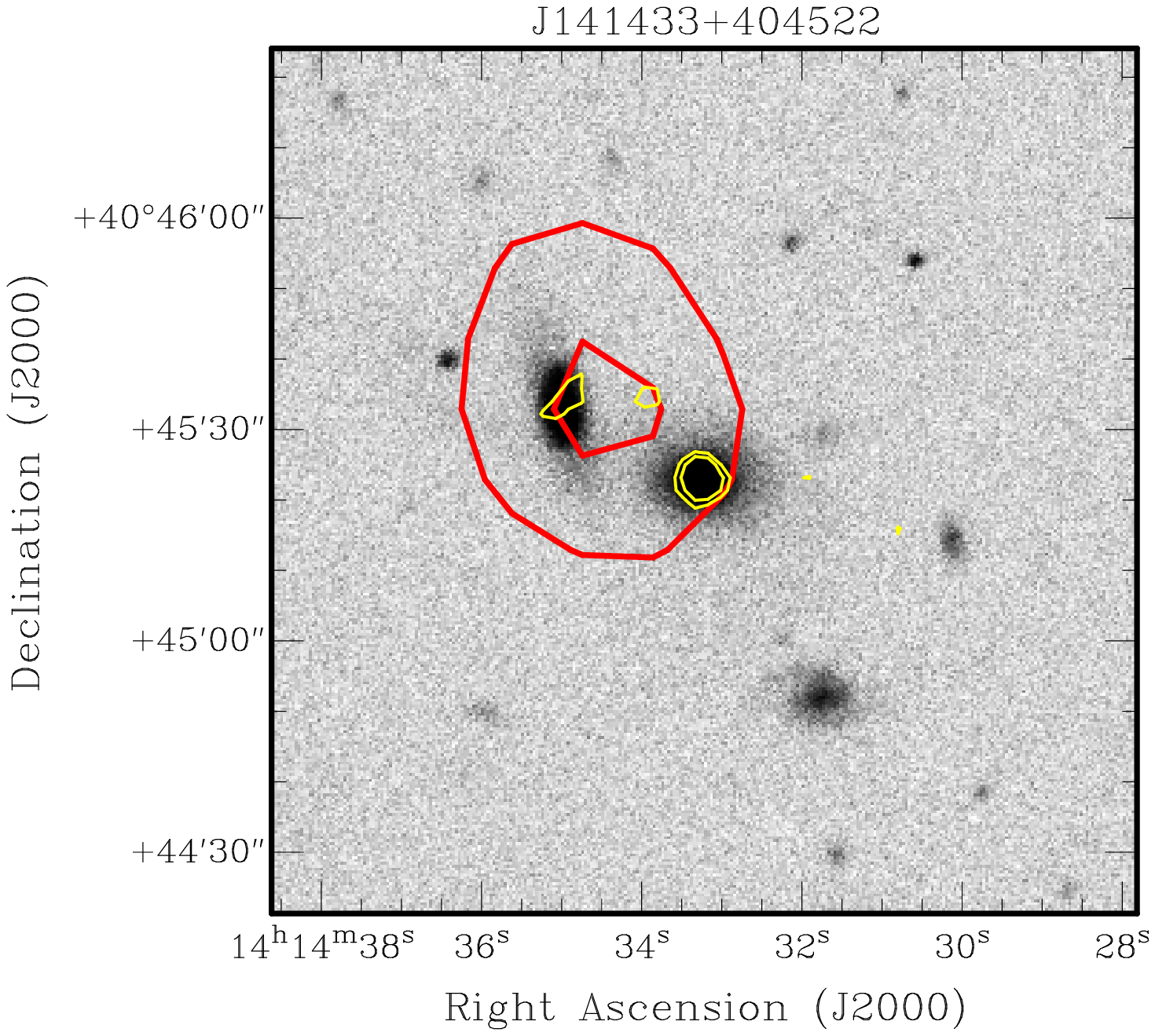}
\includegraphics[width=7.5cm,trim={2cm 3.2cm 3.8cm 3.3cm},clip]{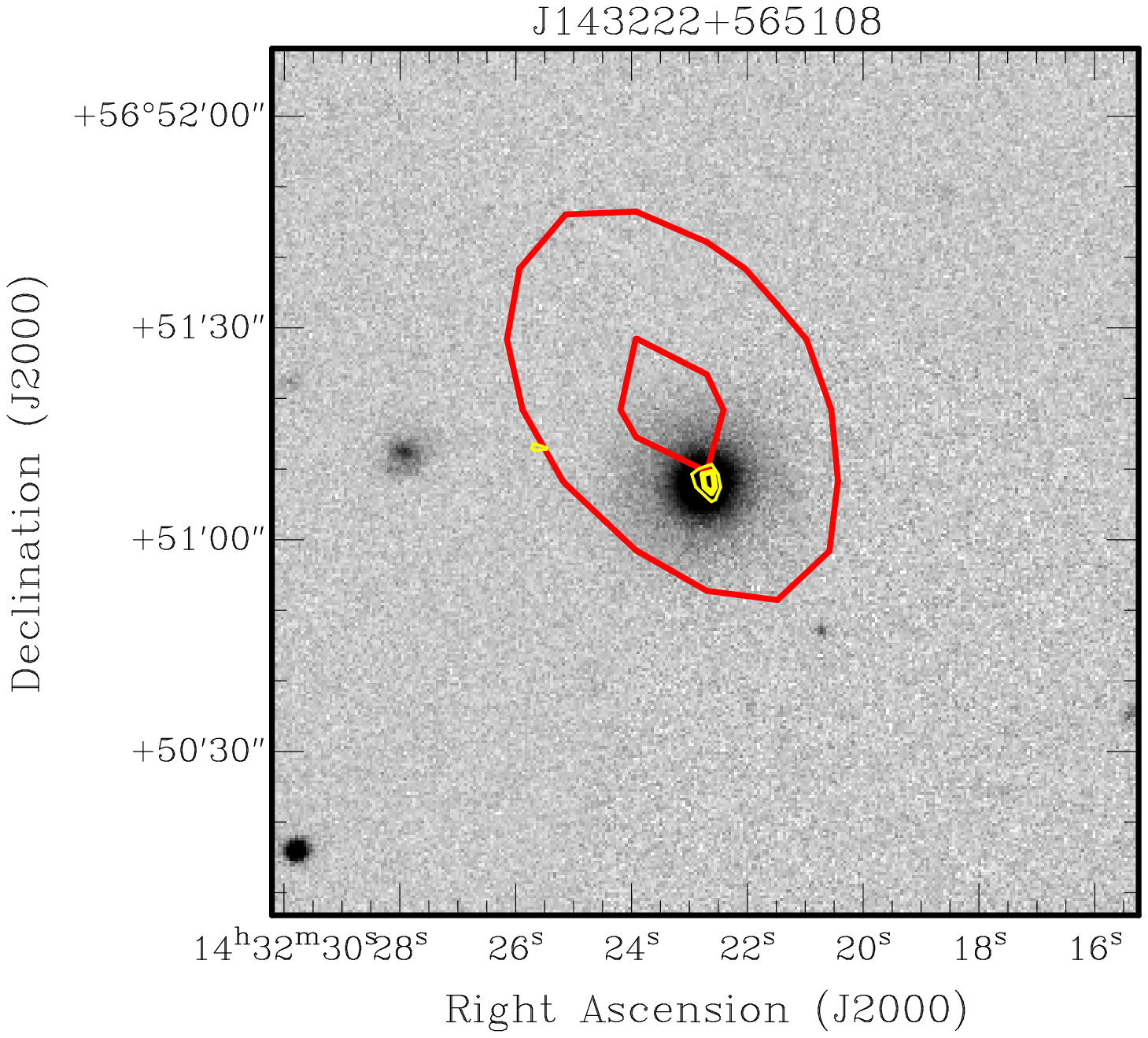}

\vspace{0.15in}
\includegraphics[width=7.5cm,trim={2cm 3.2cm 3.8cm 3.3cm},clip]{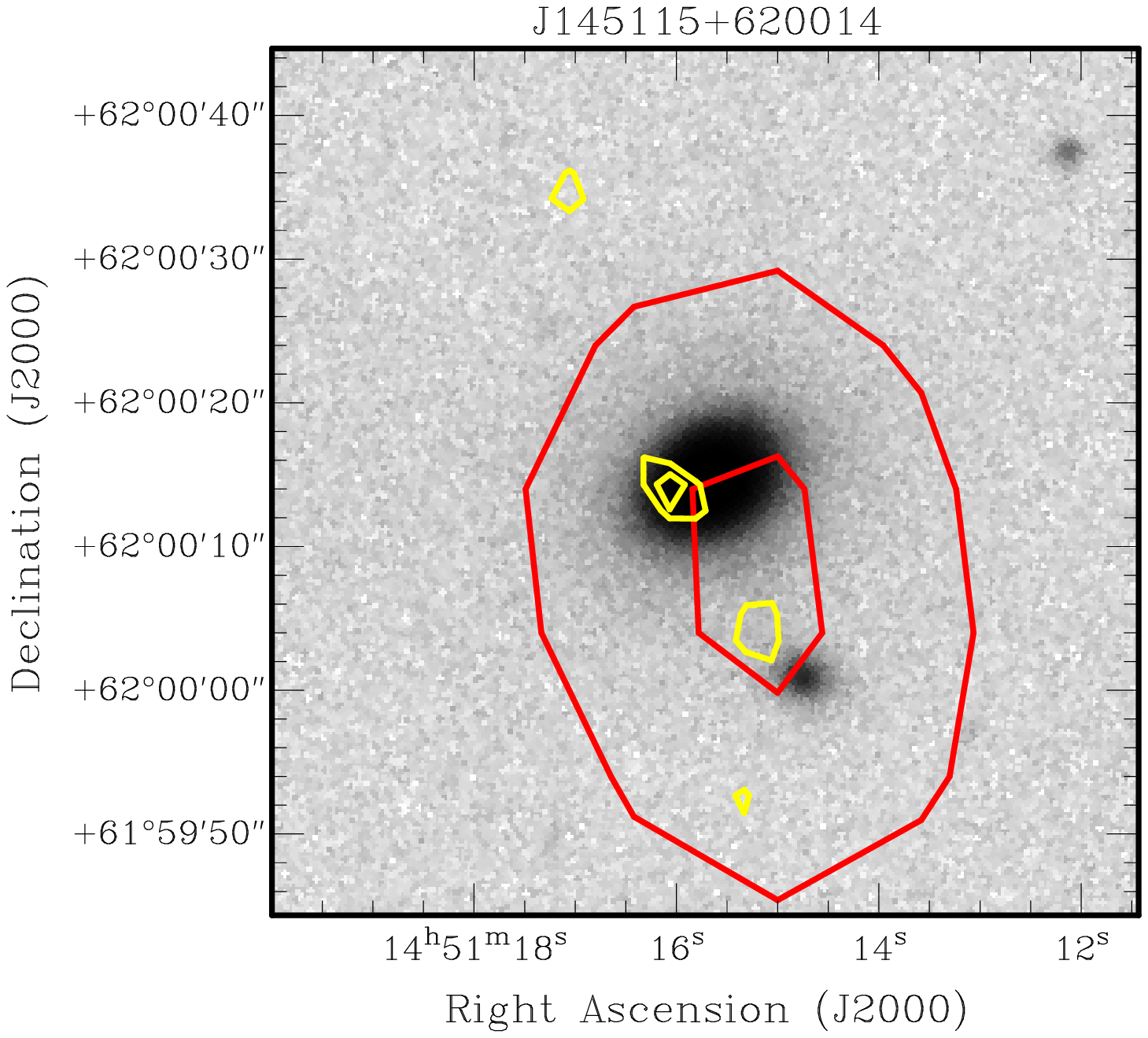}
\includegraphics[width=7.5cm,trim={2cm 3.2cm 3.8cm 3.3cm},clip]{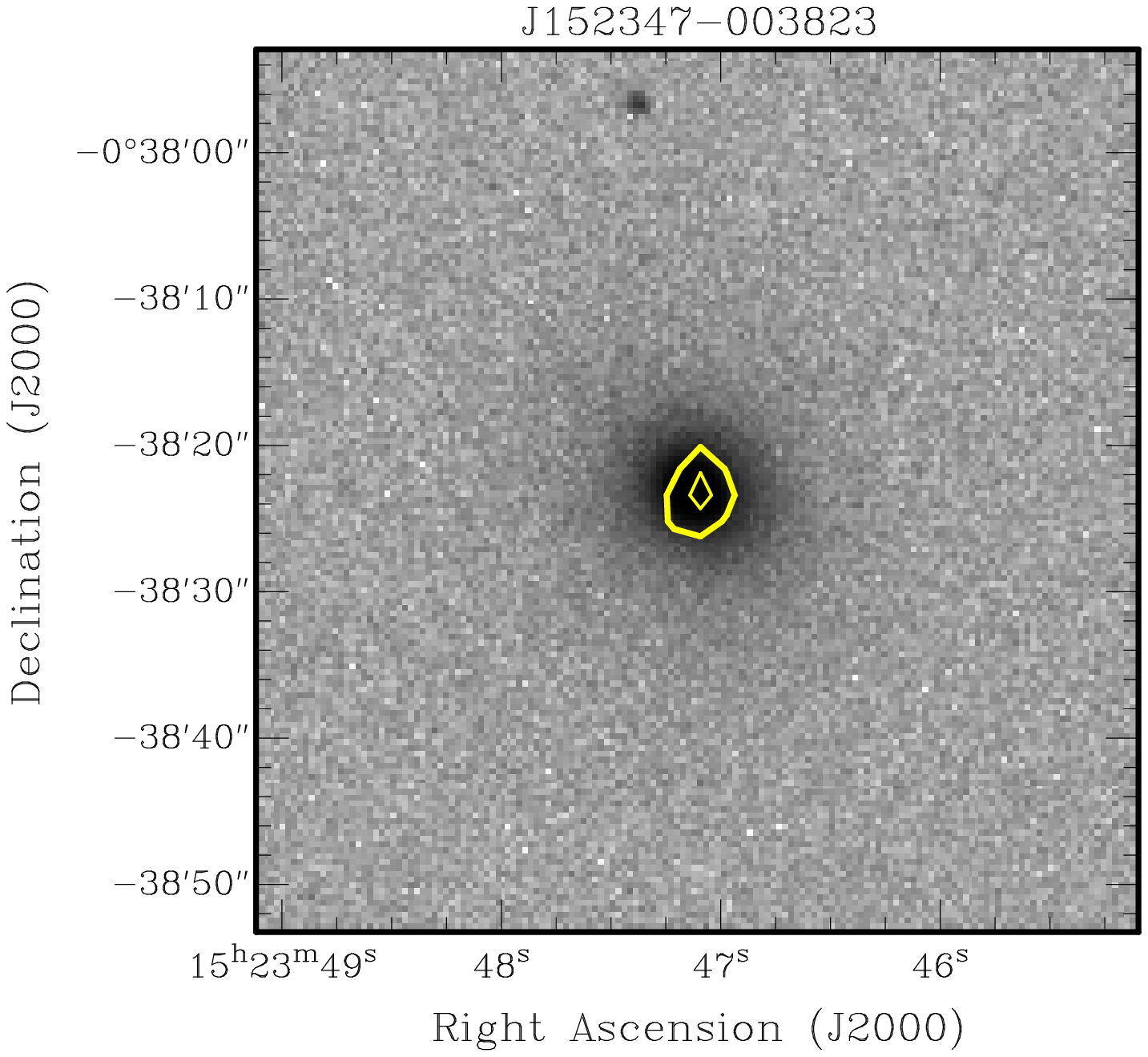}
\end{figure*}

\begin{figure*}
\centering
\includegraphics[width=7.5cm,trim={2cm 3.2cm 3.8cm 3.3cm},clip]{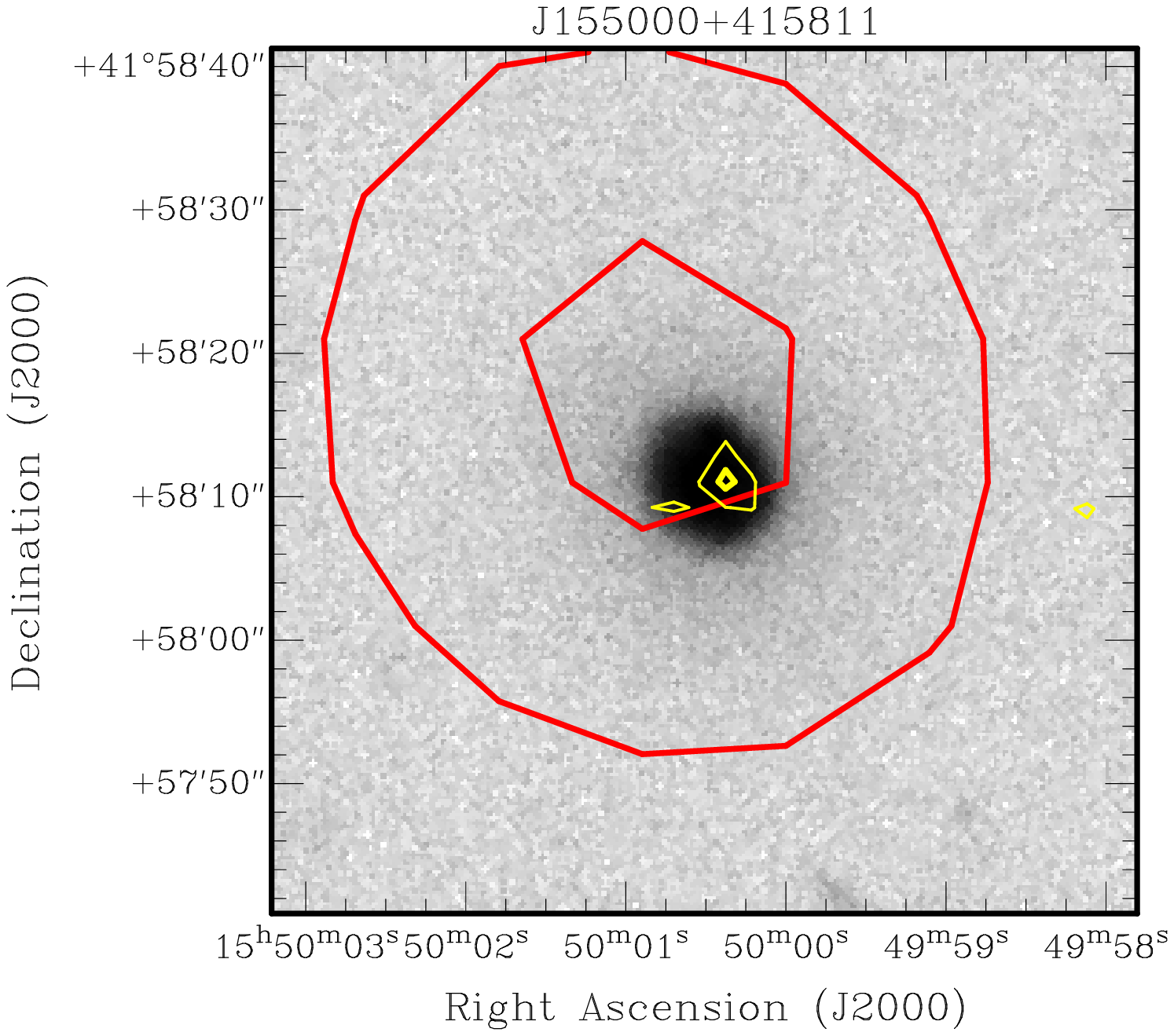}
\includegraphics[width=7.5cm,trim={2cm 3.2cm 3.8cm 3.3cm},clip]{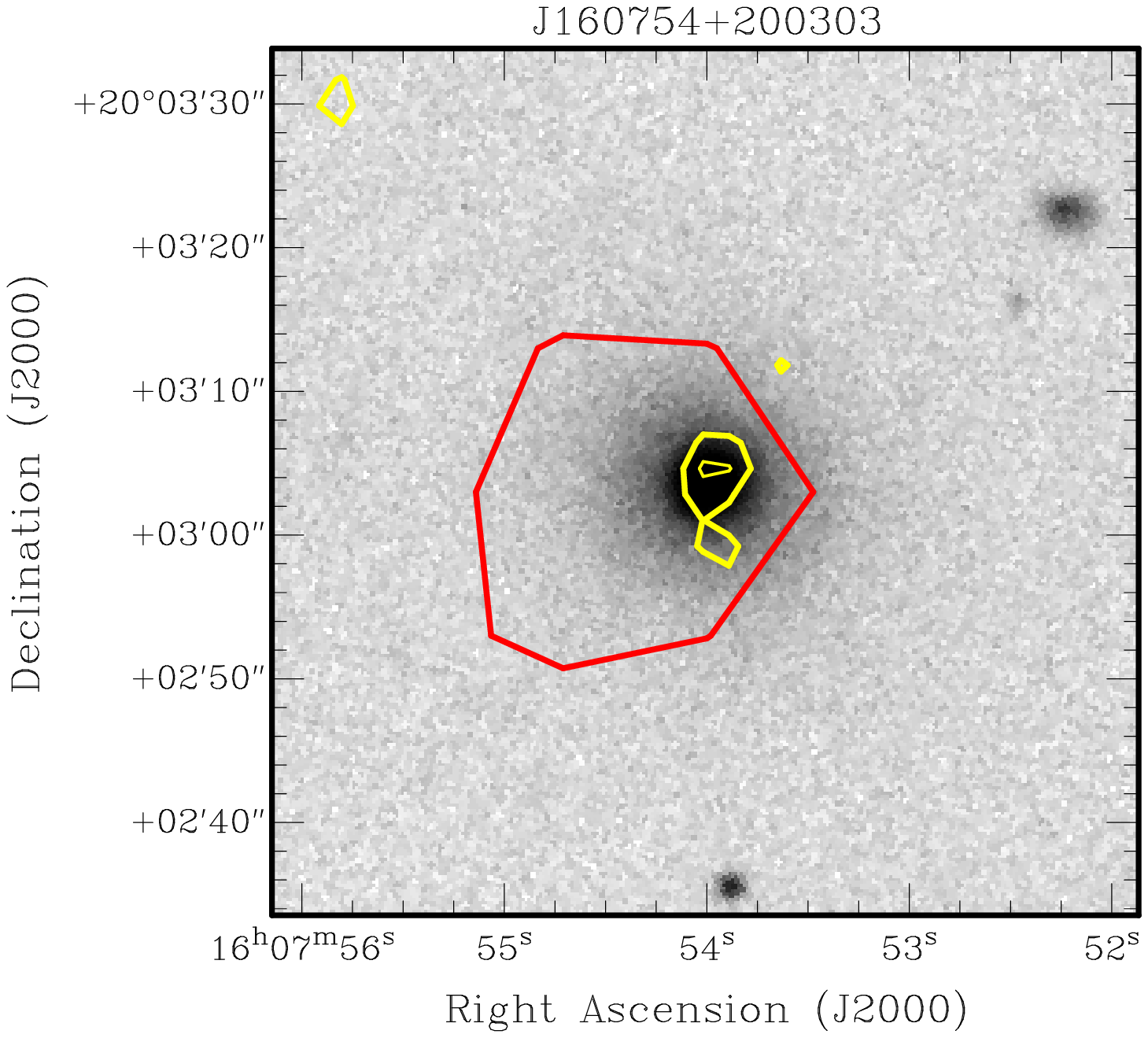}

\vspace{0.15in}
\includegraphics[width=7.5cm,trim={2cm 3.2cm 3.8cm 3.3cm},clip]{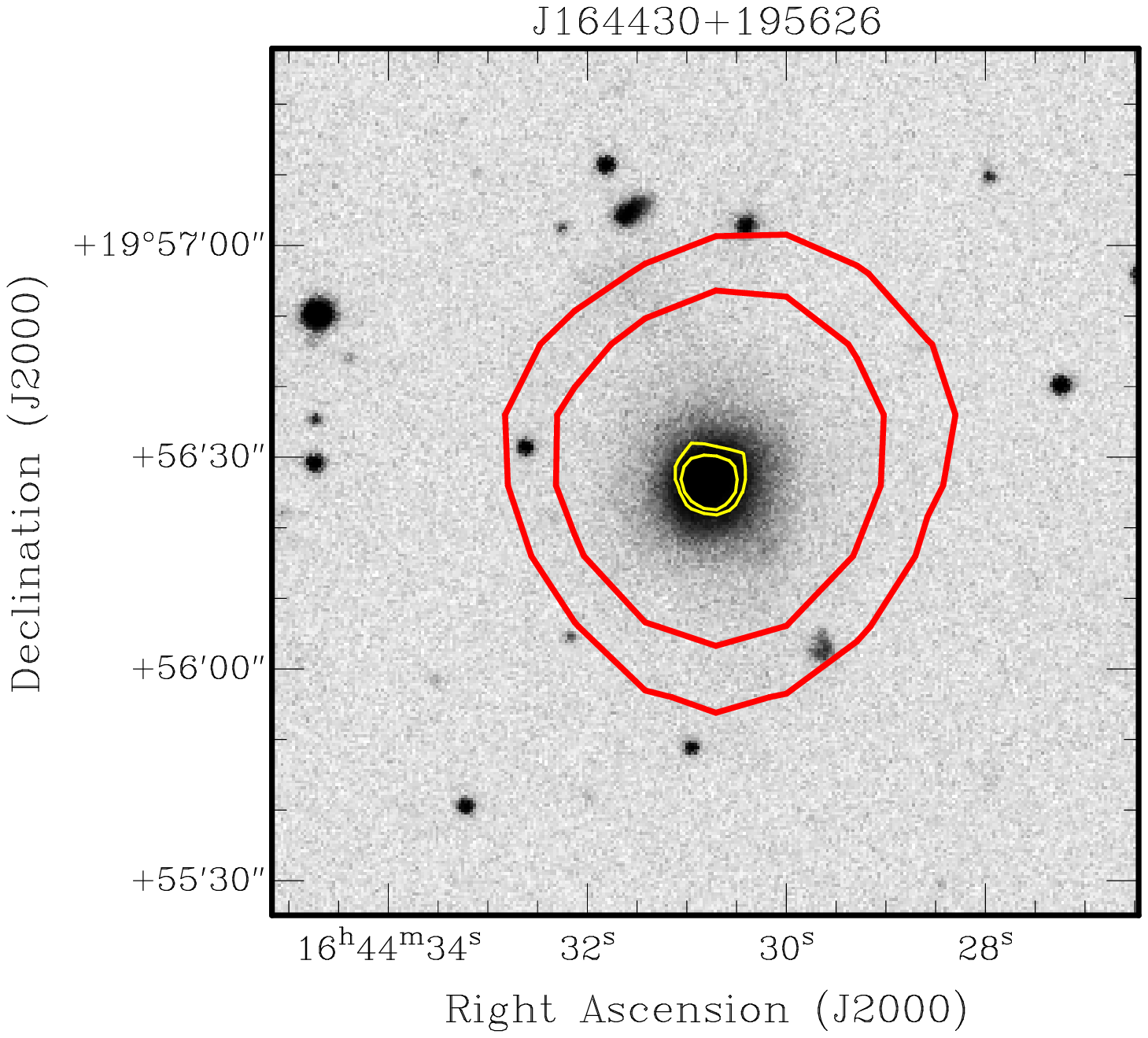}
\includegraphics[width=7.5cm,trim={2cm 3.2cm 3.8cm 3.3cm},clip]{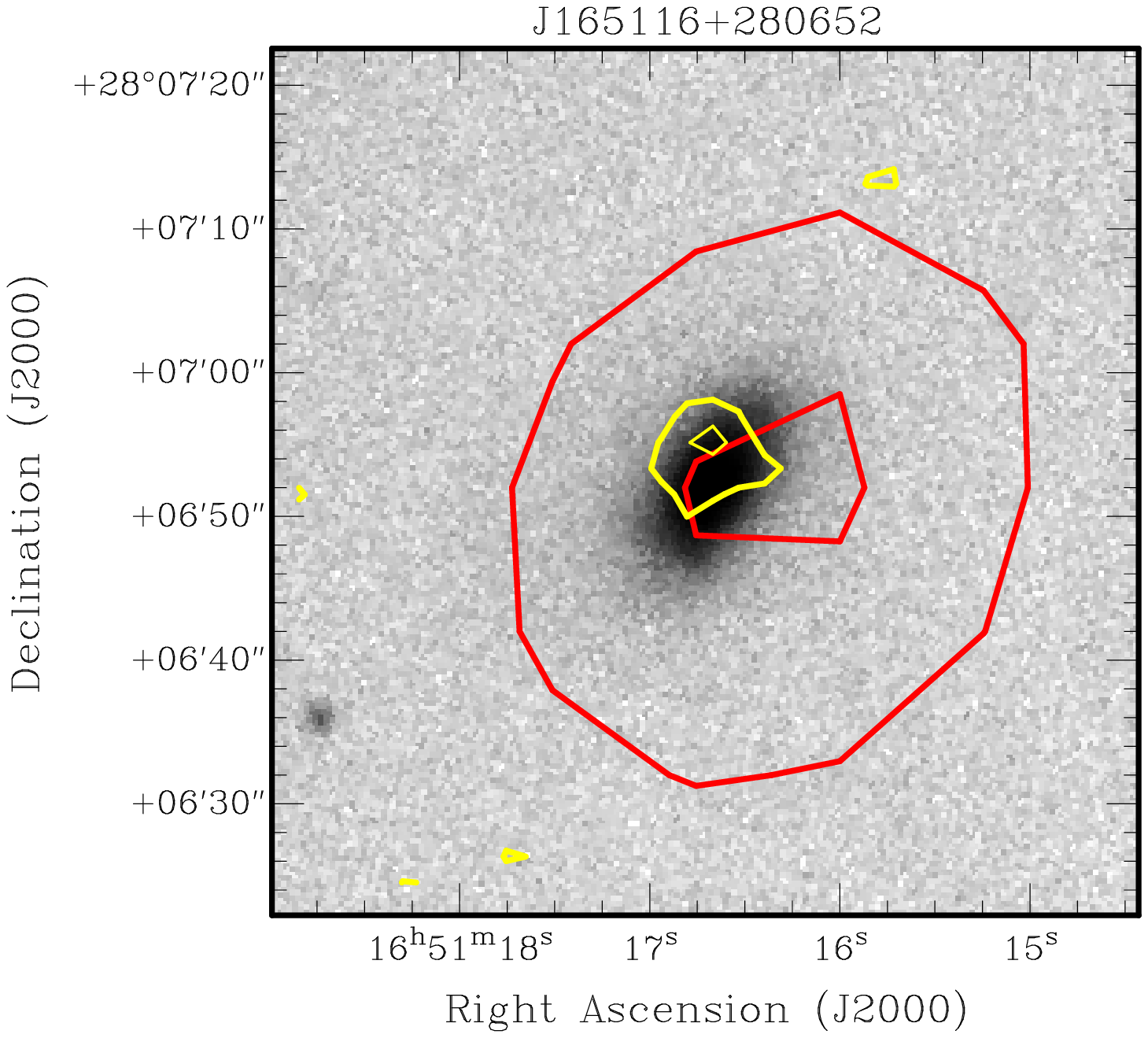}

\vspace{0.15in}
\includegraphics[width=7.5cm,trim={2cm 3.2cm 3.8cm 3.3cm},clip]{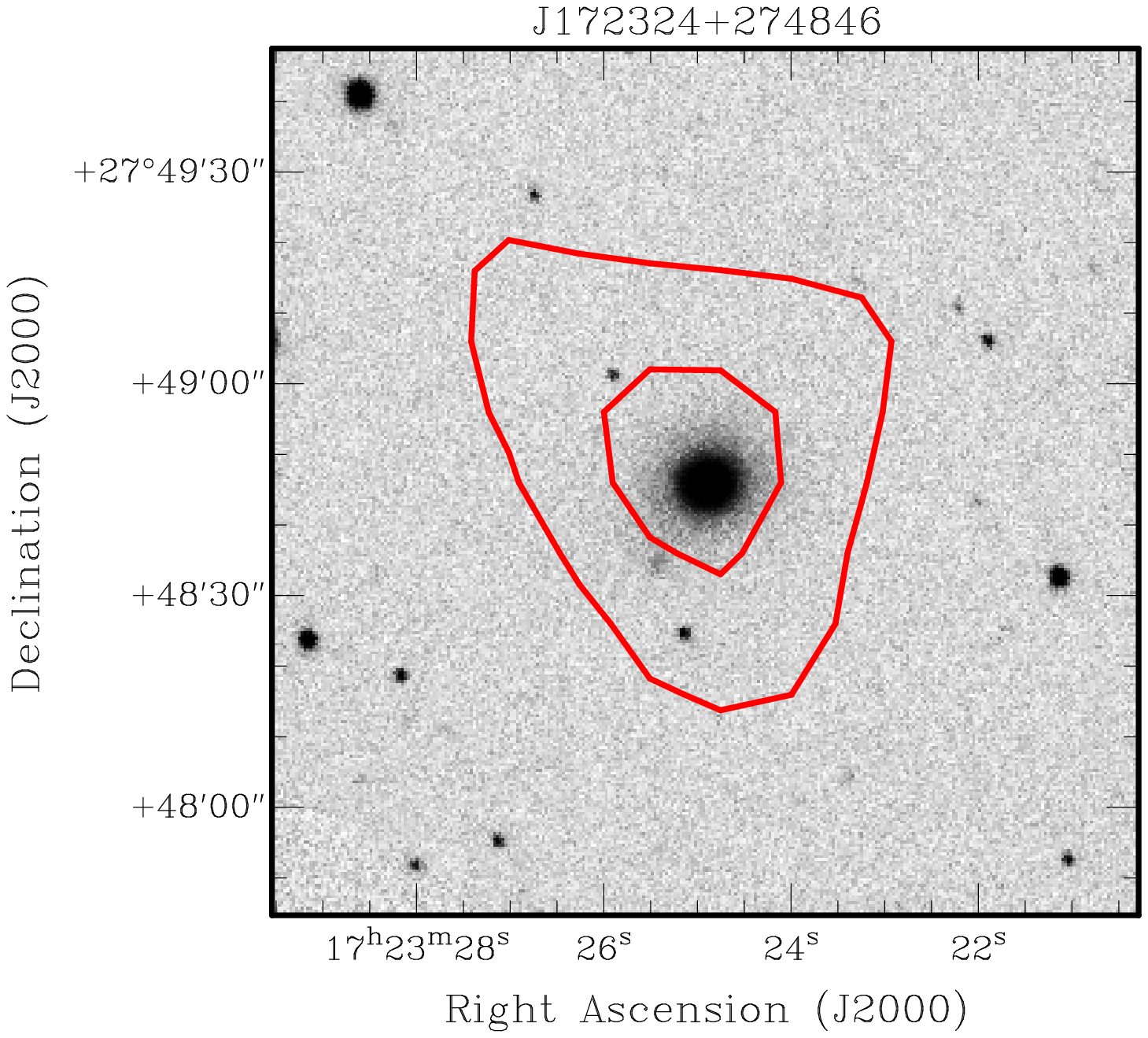}
\includegraphics[width=7.5cm,trim={2cm 3.2cm 3.8cm 3.3cm},clip]{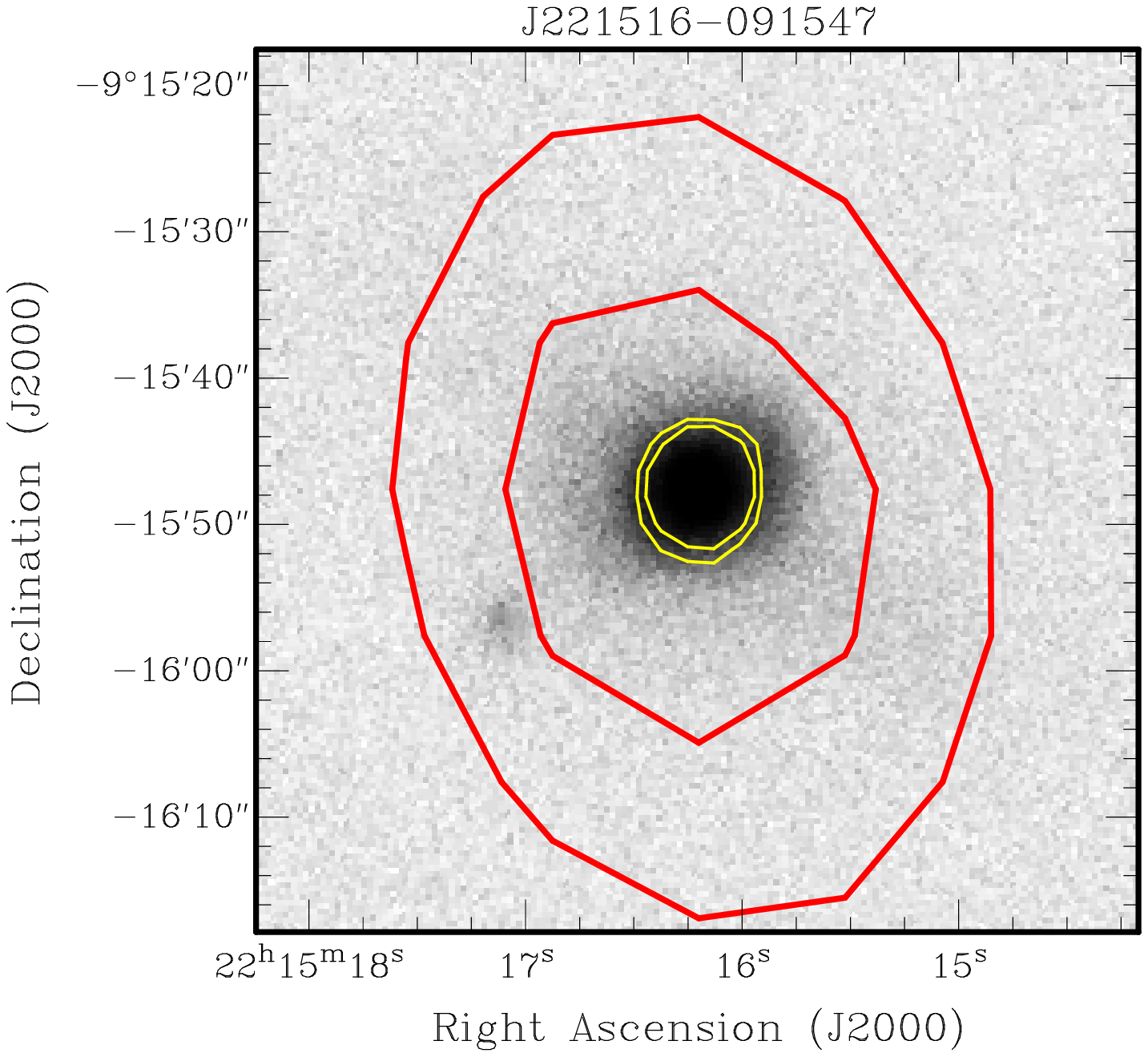}
\end{figure*}

\newpage

\bsp	
\label{lastpage}
\end{document}